\documentclass[usenatbib,useAMS]{mn2e}
\usepackage{times}
\usepackage{epsfig}

\title[Microlensing towards the SMC]
{Microlensing towards the SMC: a new analysis of OGLE\\ and EROS results}

\author[S.~Calchi~Novati et~al.]
{S.~Calchi Novati$^{1,2,3}$ 
\thanks{E-mail:novati@sa.infn.it}, 
S.~Mirzoyan$^{2,3}$, Ph.~Jetzer$^2$ and G.~Scarpetta$^{1,3,4}$\\
$^{1}$ Istituto Internazionale per gli Alti Studi Scientifici (IIASS), 
84019 Vietri Sul Mare (SA), Italy\\
$^{2}$ Institut f$\ddot{u}$r Theoretische Physik, Universit$\ddot{a}$t Z$\ddot{u}$rich, 
8057 Z$\ddot{u}$rich, Switzerland\\
$^{3}$ Dipartimento di Fisica ``E. R. Caianiello'', Universit\`a di Salerno, 
Via Giovanni Paolo II n.~132, 84084 Fisciano (SA), Italy\\
$^{4}$ Istituto Nazionale di Fisica Nucleare (INFN), Sez. di Napoli, 80126 Napoli, Italy\\
}
\begin{document}

\date{Accepted 2013 July 25.  Received 2013 July 25; in original form 2013 March 22}

\pagerange{\pageref{firstpage}--\pageref{lastpage}} \pubyear{2011}

\maketitle

\label{firstpage}

\begin{abstract}
We present a new analysis of the results of the
EROS-2, OGLE-II, and OGLE-III microlensing campaigns
towards the Small Magellanic Cloud (SMC).
Through a statistical analysis we address the issue of the
\emph{nature} of the reported microlensing candidate
events, whether to be attributed to lenses belonging
to known population (the SMC luminous components
or the Milky Way disc, to which we broadly refer
to as ``self lensing'') or to the would be population
of dark matter compact halo objects (MACHOs).
To this purpose, we present profiles of the optical depth and, comparing
to the observed quantities, we carry out analyses
of the events position and duration. 
Finally, we evaluate and study the microlensing rate.
Overall, we consider five reported microlensing events
towards the SMC (one by EROS and four by OGLE).
The analysis shows that in terms of number of events
the expected self lensing signal
may indeed explain the observed rate.
However, the characteristics of the events,
spatial distribution and duration
(and for one event, the projected velocity)
rather suggest a non-self lensing
origin for a few of them.
In particular we evaluate, through
a likelihood analysis, the resulting
upper limit for the halo mass fraction in form
of MACHOs given the expected self-lensing
and MACHO lensing signal.
At 95\% CL, the tighter upper limit, about 10\%,
is found for MACHO mass of $10^{-2}~\mathrm{M}_\odot$,
upper limit that reduces to above 20\%
for $0.5~\mathrm{M}_\odot$ MACHOs.
\end{abstract}


\section{Introduction}

The original motivation for stellar microlensing \citep{pacz86}
is the search for dark matter candidates in form of (faint) massive compact halo
objects (MACHOs) in the galactic haloes. 
Indeed, over a broad mass range of the
putative MACHO population, our current understanding of this relevant astrophysical
issue is mainly based on the results of the observational
microlensing campaigns carried out to this purpose.
On the other hand, the current understanding for the nature
of most, if not all, dark
matter at the Galactic level is from some yet
undiscovered particle \citep{strigari12}
(for a general discussion of dark matter
and gravitational lensing, we refer to \cite{bartelmann10}, \citealt{massey10}).
The probability for a microlensing event to occur
is extremely small. This is described in terms
of the microlensing \emph{optical depth} which is
of the order of $10^{-6}$ or smaller (we refer to \cite{mao12} and references therein
for an updated introduction to microlensing); therefore, dense stellar fields have
to be monitored to increase the rate of events. 
Microlensing campaigns for the search of MACHOs
have been carried out towards the Magellanic Clouds \citep{moniez10}
and M31 \citep{grg10}.
Besides the dark matter issue, meanwhile microlensing
has become an established tool for analyses
of stellar astrophysics \citep{gould01}
and, through observations towards
the Galactic bulge, for the search of extra-solar planets \citep{dominik10}. 

The Magellanic Clouds (Large and Small),
located within the Galactic halo, 
are privileged targets for the search
of microlensing events. Up to now about
20~candidate events have been reported
towards these lines of sight and important,
though not always coherent, results have been reported.
There is an agreement to exclude MACHOs as viable
dark matter candidates for masses 
below $\approx (10^{-1}-10^{-2})~\mathrm{M}_\odot$
(down to about $10^{-7}~\mathrm{M}_\odot$).
Some debate remains in the mass range
$(0.1-1)~\mathrm{M}_\odot$ 
where, according to some observational outcomes, 
a sizeable fraction, if not most of the halo mass,
may indeed be in form of compact halo objects. 
For larger values of the MACHO mass
(where the expected number of events decreases)
the limits obtained with microlensing analyses are weaker
than with other techniques \citep{yoo04b,quinn09a,quinn09b};
in this mass range it appears 
to be useful also to consider the cross-matching
of microlensing with X-ray catalogues \citep{sartore10,sartore12}.
The event duration, the ``Einstein time'' $t_\mathrm{E}$,
the main physical observable for microlensing events,
is driven by the lens mass, $m$, scaling as $\sqrt{m}$
(though it also depends on other non-directly observable
quantities as the lens-source relative transverse velocity and
the lens and source distances). 
A possibly non-exhaustive list of potential
lens populations, to which we broadly refer to
as ``self lensing'' as opposed to MACHO lensing populations, 
includes lenses belonging
to the luminous components of the Small Magellanic Cloud (SMC),
which act also as sources, and the disc of the Milky Way 
(MW; in fact, we will consider also non-luminous lenses belonging 
to these populations moving down to
the sub-stellar mass range to include also brown dwarfs).
The suggestion that the events observed towards the Magellanic
Clouds may not be due to MACHOs dates back at least to
the analyses of \cite{sahu94}, \cite{wu94}, \cite{gould95} and has been
thereafter the object of several analyses \citep{salati99,distefano00,evans00,gyuk00,jetzer02}.
It is therefore relevant to reliably determine
the signal expected from self-lensing lens populations
as compared to that of MACHO lensing.

More specifically, the MACHO collaboration claimed for a mass halo fraction
in form of $\sim 0.5~\mathrm{M}_\odot$ MACHOs of about $f\sim 20\%$ out 
of observations of 13-17 candidate microlensing
events towards the Large Magellanic Cloud (LMC) \citep{macho00}, a result
further discussed in \cite{bennett05}
where in particular the microlensing nature of
10-12 out of the original set of 13 candidate events
has been confirmed. On the other hand, in disagreement with this result,
the analyses of the EROS \citep{eros07},  
and the OGLE collaborations, for both OGLE-II \citep{ogle09,ogle10} and OGLE-III \citep{ogle11,ogle11b},
out of observations towards both the LMC and SMC, concluded 
by putting extremely severe \emph{upper} limits
on the MACHO contribution also in this mass range.
In particular, at 95\% CL, the EROS collaboration reported
an upper limit $f=8\%$ for $0.4~\textrm{M}_\odot$ MACHOs, and
OGLE $f=6\%$ for $0.4~\textrm{M}_\odot$ MACHOs 
and $f=4\%$ in the mass range between $0.01$ and $0.15~\textrm{M}_\odot$.

Rather than addressing, as we also mainly do in this work,
the issue on the \emph{lens}  nature, whether self lensing or MACHO lensing,
one may also consider different \emph{source} populations
which may possibly enhance the microlensing rate
(see for instance \cite{rest05} and
reference therein for a broad overall discussion
of the different possible source and lens populations).
For the specific case of the LMC, 
recently \cite{besla13} proposed, as possible 
sources, a SMC stripped population 
(still to be observed, though) lying \emph{behind} the LMC,
which may explain simultaneously  both the MACHO 
and the OGLE observational results towards the LMC
(see however \cite{nelson09} who, on a general ground,
concluded against the possibility for the sources
to lie behind the LMC).

With respect to the LMC, the case of the SMC is somewhat peculiar.
As further discussed below, the SMC is quite elongated
along the line of sight. As the microlensing cross-section,
the Einstein radius, is proportional to the (square root of)
the source-lens distance, an elongated structure is expected to enhance
the SMC self-lensing signal. As a result, the ratio of self lensing
versus MACHO lensing (if any) is larger than towards the
LMC making overall more difficult to disentangle the two signals
and to draw stringent conclusions on the issue of MACHOs.
On the other hand, the characteristics that differentiate
the two lines of sight can be considered as a strength
when cross-matching the results.

In previous analyses, we have addressed the issue
of the nature of the reported events towards the LMC
by the MACHO \citep{mancini04,novati06}
and the OGLE collaborations \citep{novati09b,novati11}.
In this paper, we report  a detailed analysis of the 
EROS and the OGLE observational campaigns 
towards the SMC (as further discussed
below, we do not include data from the observational
campaign carried out by the MACHO collaboration along this line of sight).
The underlying idea behind our approach is
to  characterize statistically,
starting from a reliable model
for all possible lens populations,
the observed versus the expected signal in order to address the issue of the 
nature of the reported microlensing candidate events.
First, we evaluate profiles of the optical depth, in particular for SMC self lensing.
This tells us how the SMC structure
is reflected in the expected microlensing signal
and carries information  on the overall spatial density 
of the given lens population.
To include within the analysis the specific
characteristics of the observed events,
in particular number, position and duration,
we carry out an investigation based on the microlensing rate
which then allows  us to derive limits on the halo mass fraction in form of MACHOs.

The plan of the paper is as follows. 
In Section~\ref{sec:model},  we describe the models used in our analysis,
with a particular attention to the SMC structure.
In Section~\ref{sec:obs}, we resume 
the status and the results of previous and ongoing microlensing campaign towards the SMC.
In Section~\ref{sec:analysis}, we present our analysis.
In Section~\ref{sec:tau}, we present the profiles of the optical depth.
In Section~\ref{sec:rate}, we introduce the microlensing rate, our main tool of investigation.
In Section~\ref{sec:res}, we derive the expected microlensing quantities, number of events and duration.
In Section~\ref{sec:nat}, 
we address the issue of the possible nature of the reported
observed events and in particular we evaluate the limits on dark matter in form of compact halo objects.
In Section~\ref{sec:compare}, we compare our results to previous ones
towards the SMC and critically analyse, as for the search of MACHOs, the line of sight
towards the SMC against that towards the LMC.
Finally, in Section~\ref{sec:end}, we present our conclusions.

\section{Model} \label{sec:model}

The microlensing quantities, the microlensing optical depth and the microlensing rate,
depend on the underlying astrophysical model. In particular, the optical depth
depends uniquely on the lens (and source) population
spatial density, whereas the microlensing rate depends also on the lens mass function and 
the lens-source relative velocity. Indeed, for the more common situation of
a point-like single lens and source with uniform relative motion, the only physical observable
characterizing the events is the Einstein time, $t_\mathrm{E}$,
which is a function of the lens mass, the lens-source relative velocity and
the lens and source distances. None of these quantities, however, is directly observable.
The underlying astrophysical models are therefore essential to assess the characteristics
of the expected signal from all the possible lens populations.
In the present case: self lensing, which fixes the background level,
and MACHO lensing, the ``signal'' we want to constraint.

In the following analysis we consider, as possible
lens populations, SMC and MW stars (and brown dwarfs),
both contributing to the self-lensing signal,
and the would be population of compact halo objects
in the MW halo, which we describe in turn.

\subsection{The SMC: structure and kinematics}\label{sec:smc}

\subsubsection{Structure} \label{sec:structure}
The SMC is a dwarf irregular galaxy orbiting
the MW in tight interaction with the (larger) LMC 
\citep{vdb99,mcconnachie12}. Also because
of this complicated dynamical situation, the detailed spatial structure
and overall characteristics of the SMC are still debated.
For the overall SMC \emph{stellar} mass, which is a quantity
of primary importance to our purposes, 
being in the end (almost) directly proportional
to the SMC optical depth and number of expected events,
we use $M_* = 1.0\times 10^9~\mathrm{M}_\odot$ (within $5~\mathrm{kpc}$
of the SMC centre), which is the value of the ``fiducial'' 
model of \cite{bekki09b} (see also \citealt{yoshizawa03}).
According to reported values of the SMC luminosity, this correspond roughly
to a mass-to-light ratio within the range $M/L_V\sim 2-3$.
We recall that \cite{mcconnachie12} reports $M_*=4.6\times 10^8~\mathrm{M}_\odot$,
which is half smaller than our fiducial value,
and that \cite{bekki09a}, for a stellar luminosity $4.3\times 10^8~\mathrm{L}_\odot$,
consider the ``reasonable'' range for the mass-to-light ratio
to be $M/L_V\sim 2-4$, depending in particular on the fraction
of the old stellar population. In a previous work, \cite{stanimirovic04},
for a stellar luminosity $3.1\times 10^8~\mathrm{L}_\odot$,
estimated a total stellar mass of the SMC $1.8\times 10^9~\mathrm{M}_\odot$
(within $3~\mathrm{kpc}$ of the SMC centre). 
Overall, the systematic uncertainty on this relevant quantity 
we find in literature is of about a factor of 2. 

The total \emph{dynamical} mass of the SMC has also been the object of several investigations. 
\cite{stanimirovic04} report  $2.4\times 10^9~\mathrm{M}_\odot$
within $3~\mathrm{kpc}$, a result confirmed in \cite{harris06} who report values in the range
$1.4-1.9\times 10^9~\mathrm{M}_\odot$ within $1.6~\mathrm{kpc}$
and a less well constrained mass within $3~\mathrm{kpc}$
between 2.7 and $5.1\times 10^9~\mathrm{M}_\odot$.
These values therefore suggest the existence of a dark matter
component even in the innermost SMC region, as thoroughly 
discussed in \cite{bekki09a}. 

According to the star formation history of the SMC \citep{harris04}
we can broadly distinguish two components: an old star (OS)
and a young star (YS) population. Several analyses
have shown that indeed, besides their age, these
populations also show different morphology structures.
We base our analysis upon the recent work of \cite{grebel12},
which in turn is based on the OGLE-III SMC variable stars data
but see also, among others, \cite{kapakos11,kapakos12,nidever11,subsub09,subsub12}.
In particular, \cite{grebel12} address the issue
of the three dimensional SMC structure
based on the analysis of RR Lyr{\ae} stars and
Cepheids as tracers of the old and young populations,
respectively. \cite{grebel12} report estimates
for the position and the inclination angles
and for the line-of-sight depth which 
is a crucial quantity for microlensing purposes
as the microlensing cross-section, and in the end
the microlensing rate, grows with the lens-source
distance, so that a large SMC intrinsic depth
enhances the SMC self-lensing signal
whereas, on the other hand, the details of the inner SMC structure
are not essential to determine the expected
lensing signal for the MW lens populations (Section~\ref{sec:tau}).
Moreover, \cite{grebel12}  show contour plots for
the stellar density of RR Lyr{\ae} stars and
Cepheids not only on the plane of the sky but also
on the distance-declination and the distance-right ascension
planes. For fixed values of the position and inclination
angles and line-of-sight depth we therefore
build our model trying to broadly match
this full three-dimensional view of the SMC.

As a model for both populations we choose a spheroidal structure
with a fully Gaussian profile for the YS population
\begin{equation} \label{eq:ys}
\rho_\mathrm{SMC}^{\mathrm(YS)} = \rho_0^{\mathrm(YS)}\, \exp\left[-\frac{1}{2}
\left(\left(\frac{\xi}{\sigma_\xi}\right)^2+
\left(\frac{\eta}{\sigma_\eta}\right)^2+\left(\frac{\zeta}{\sigma_\zeta}\right)^2\right)\right]\,.
\end{equation}
For the OS population, we keep the Gaussian profile along the line of sight
(which in particular ensures a roughly constant
line-of-sight depth), and a smoother exponential profile
in the orthogonal plane
\begin{equation} \label{eq:smc}
\rho_\mathrm{SMC}^{\mathrm(OS)} = \rho_0^{\mathrm(OS)}\, \exp\left[-\sqrt{\left(\frac{\Xi}{\Xi_0}\right)^2+
\left(\frac{\Upsilon}{\Upsilon_0}\right)^2}\right] 
\exp\left[-\frac{1}{2}\left(\frac{Z}{\sigma_Z}\right)^2\right]\,.
\end{equation}
With respect to the north direction, the value of the position angle is 
fixed at $66^\circ$ and $83^\circ$ for the YS
and OS populations, respectively. The YS are strongly inclined
by an angle of $74^\circ$ with the north-east part nearer to us.
The OS, on the other hand, do not show an inclination
significantly different from zero. We assume therefore a zero inclination angle
also in agreement with the analysis of \cite{subsub12}.
The line-of-sight depth is $4.2~\mathrm{kpc}$ and
in the range $5.4-6.2~\mathrm{kpc}$ for the old
and young populations, respectively. These values
are as reported in the analysis of \cite{grebel12}
to which we also refer for a critical discussion
of previous analyses.
The reference frames $(\xi,\eta,\zeta)$ and $(\Xi,\Upsilon,Z)$
are directed along the principal axes of the YS and OS spheroid, respectively.
For the YS population we fix $(\sigma_\xi,\sigma_\eta,\sigma_\zeta) 
= (0.8,3.5,1.3)~\mathrm{kpc}$, for an overall elongated bar-shape.
Starting  from the $x,y,z$ frame, Fig.~\ref{fig:campi} and
with the $z$ axis going from the SMC centre to the observer,
we move to the $\xi,\eta,\zeta$ principal axes frame
through a counterclockwise 
rotation around the $z$ axis of the position angle
followed by a  counterclockwise rotation around the new $\xi$ axis
of the inclination angle.  
For the OS population, $\sigma_Z=2.1~\mathrm{kpc}$
and $(\Xi_0,\Upsilon_0)=(0.8,1.2)~\mathrm{kpc}$,
with the reference frame $(\Xi,\Upsilon,Z=z)$ obtained,
from the $x,y,z$ one, through a counterclockwise 
rotation around the $z$ axis of the position angle.
Following again the analysis of \cite{bekki09b}
we assume a OS over YS mass-ratio of 6:4,
as in their fiducial model
(this quantity is however not well
constrained by the simulation 
and overall its estimate is still not robust).
Accordingly, the central density
values are fixed to $3.9$ and $8.5\times 10^6~\mathrm{M_\odot}
\mathrm{kpc}^{-3}$ for the old and young star
population, respectively. 

The centre and the distance of the SMC
are both not too well constrained
and in particular an offset of the young and old population,
both in distance and in position, which may indeed be relevant
for the evaluation of the microlensing quantities,
has been discussed by several authors.
Here again we follow the analysis of  \cite{grebel12},
and reference therein, and assume,
for our fiducial model,
the same centre and distance for both
populations. In particular, we choose 
the optical centre reported by \cite{gonidakis09} 
$\alpha=0^\mathrm{h} 51^\mathrm{m}$
and $\delta=-73^\circ.1$ (J2000)
and a distance to the SMC of $61.5~\mathrm{kpc}$,
the median distance of RR Lyr{\ae} stars \citep{grebel12} 
found in agreement with that of the Cepheids.
Finally, we fix the tidal radius
of the SMC at $12~\mathrm{kpc}$.

Because of a relative shift in distance between
the OS and YS populations
is still compatible with the data and
may be expected to enhance the microlensing rate,
as a test model we consider the case where
the centre of the YS population is shifted by $2~\mathrm{kpc}$
behind that of the OS one, at $63.5~\mathrm{kpc}$,
rescaling (increasing) the YS axes ratio to keep the same
shape on the plane of the sky and changing accordingly
the central normalization.

In \cite{eros98}, the EROS collaboration introduced
an SMC model for an estimate of the SMC self-lensing
optical depth which has therefore become an
often quoted ``fiducial'' value for this quantity.
The SMC, for a  total stellar  mass
of $\sim 1\times 10^9~\mathrm{M}_\odot$ (a value 
that matches the one we use in our model),
is approximated with a single population prolate ellipsoid
elongated along the line of sight with exponential
profile. The radial scale length, transverse to the line
of sight, is fixed at $0.54~\mathrm{kpc}$ and the scale height
along the line of sight is left free to vary in the range
$2.5-7.5~\mathrm{kpc}$. We recall that
the scale height is smaller than
the depth by a factor 0.4648 \citep{grebel12},
so that $2.5~\mathrm{kpc}$ is the value that
better matches our ones. 
Although clearly disfavoured, in view of the more recent
pieces of observational evidence, 
we consider useful to compare to this model in consideration
of its importance in the microlensing literature.
Indeed, being peculiarly different
but still characterized by the same overall quantities (in particular,
stellar mass and scale height), it represents
a useful test case against 
our fiducial model\footnote{There is a caveat concerning
the total mass and the corresponding normalization of this model. In fact, 
although \cite{eros98} report a stellar mass of
$\sim 1\times 10^9~\mathrm{M}_\odot$,
which corresponds to our chosen normalization
for the SMC luminous mass within $5~\mathrm{kpc}$,
we have to introduce a multiplicative factor 1.6, which we use, in the density
normalization, with respect
to the values reported in \cite{eros98}, 
to match the overall mass of our model
within the tidal radius.
}.

\subsubsection{Kinematics}
We consider the velocity of SMC lenses as due to
the sum of a non-dispersive component and a
dispersive component. For the systemic proper motion,
we follow the analysis of \cite{kalliva06c}
with $(\mu_W,\mu_N)=(-1.16,-1.17)~\mathrm{mas~yr}^{-1}$
(in acceptable agreement with the outcome of
the analysis of \citealt{piatek08}),
with an observed line-of-sight velocity $146~\mathrm{km~s}^{-1}$
\citep{harris06}. For the YS population,
we also introduce a solid body rotation around the $\xi$ axis 
(Section~\ref{sec:structure}) linearly
increasing up to $60~\mathrm{km~s}^{-1}$ with turnover radius at $3~\mathrm{kpc}$
\citep{stanimirovic04}. For the dispersive velocity component,
we assume an isotropic Gaussian distribution
(\cite{harris06} report the line-of-sight velocity distribution
to be well characterized by a Gaussian with a velocity dispersion
profile independent from the position). For the velocity
dispersion values we, again, follow those of the fiducial model
of \cite{bekki09b}, with $\sigma=30$
and $20~\mathrm{km~s}^{-1}$ for the old and young star populations, respectively.
This is in good agreement with $\sigma=27.5~\mathrm{km~s}^{-1}$
for the old populations stars analysed in \cite{harris06}
and with the analysis of \cite{evans08}.

\subsection{The MW disc and dark matter halo}\label{sec:mw}

For the MW disc, with assumed distance from the Galactic Centre 
$8~\mathrm{kpc}$ and local circular velocity $220~\mathrm{km~s}^{-1}$ 
(in agreement, for instance, with the recent analysis of \citealt{bovy12}),
we closely follow the analysis in \cite{novati11}
with double exponential profiles thin and thick disc components
with, respectively, local density $0.044~(0.0050)~\mathrm{M}_\odot \mathrm{pc}^{-3}$,
scale height $250~(750)~\mathrm{pc}$, scale length $2.75~(4.1)~\mathrm{kpc}$
\citep{kroupa07,juric08,dejong10}, and line-of-sight dispersion of $30~(40)~\mathrm{km~s}^{-1}$.

For the Galactic dark matter halo, in order to coherently compare
with previous microlensing analyses, we assume the ``standard'' 
\cite{macho00} pseudo-isothermal spherical density profile with core radius 
$5~\mathrm{kpc}$ \citep{deboer05,weber10} and \cite{macho00} central density 
$0.0079~\mathrm{M}_\odot \mathrm{pc}^{-3}$
(in excellent agreement with up-to-date estimates as in
\cite{tremaine12}, see however \citealt{garbari12}) 
for an isotropic Gaussian distribution velocity
with line-of-sight dispersion  $155~\mathrm{km~s}^{-1}$.

\subsection{Mass function}\label{sec:imf}

For all the self-lensing populations, we assume
a power-law mass function ($\mathrm{d}N/\mathrm{d}M\propto M^{-\alpha}$).
For  MW disc lenses, following \cite{kroupa11},
we assume slopes 1.3 and 2.3 
in the mass ranges $(0.08-0.5),\, (0.5-1)~\mathrm{M}_\odot$,
and a present-day mass function slope 4.5
above $1~\mathrm{M}_\odot$. As un upper limit for the
stellar mass, which we use to normalize the 
mass function to match the value of the mass of the
lens population \citep{jetzer02}, we use $120~\mathrm{M}_\odot$.
For the SMC YS population, we make use
of the results presented in \cite{kalirai13}
where the initial mass function of the SMC 
is evaluated using ultradeep \emph{Hubble Space Telescope}
imaging in the outskirts of the SMC (along the line
of sight of the foreground globular cluster 47 Tuc).
In particular, \cite{kalirai13} conclude for a mass function
well represented by a single power-law form
with slope $1.90^{+0.15}_{-0.10}$ in the mass range
$(0.37-0.93)\,\mathrm{M}_\odot$, with also indications
of a turnover at the low-mass end which would
not be reproduced by the same power-law index.
Finally, we extend the \cite{kalirai13} mass function
in the mass range $(0.37-1.00)\,\mathrm{M}_\odot$
and use the disc \cite{kroupa11} otherwise.
The upper limit of integration for the lens mass within the 
evaluation of the microlensing rate 
is in principle set so to avoid any possible ``visible'' lens.
As for SMC lenses, above $1~\mathrm{M}_\odot$, however,
because of the extreme steepness of the mass
function in this mass range, the exact value
turns out to be in fact irrelevant
in view of the calculation of the relevant microlensing
quantities (expected number and duration of events). 
To be fixed, we choose for this parameter
the value $2~\mathrm{M}_\odot$.
The microlensing results are on the other hand more sensitive 
to the value of this threshold  for possible nearby MW lenses,
given also the rapid variation of
the mass-magnitude relationship with the lens distance.
As a very conservative choice, intended to provide
an upper limit for the expected microlensing
signal of this population, we set the same value as for SMC lenses
with the caveat that, reducing this threshold
to $1~\mathrm{M}_\odot$
the related expected signal would reduce
of about 10\%. Given that (Section~\ref{sec:res})
the expected MW signal represents about
10\%-20\% of that of SMC lenses, 
the impact of this choice remains small. 
For the SMC OS population we use,
following the results obtained for the Galactic bulge \citep{zoccali00},
a power law with slope 1.33 
in the range $(0.08-1)~\mathrm{M}_\odot$,
with upper limit for integration and normalization 
also fixed at $1~\mathrm{M}_\odot$.
Besides the MW and SMC stellar populations,
we also include a \emph{brown dwarf} component, in the mass range
$0.01-0.08~\mathrm{M}_\odot$ with power-law mass function index $0.3$
\citep{allen05,kroupa11}. Following the local analysis of \cite{chabrier03},
we attribute to this component 5\% of the overall relative
stellar mass component.

For dark matter halo lenses we test a series
of delta mass function in the mass range
$10^{-5}-10^2~\mathrm{M}_\odot$. 

\section{Microlensing towards the SMC:
The EROS and the OGLE campaigns}\label{sec:obs}
\begin{figure}
\includegraphics[width=84mm]{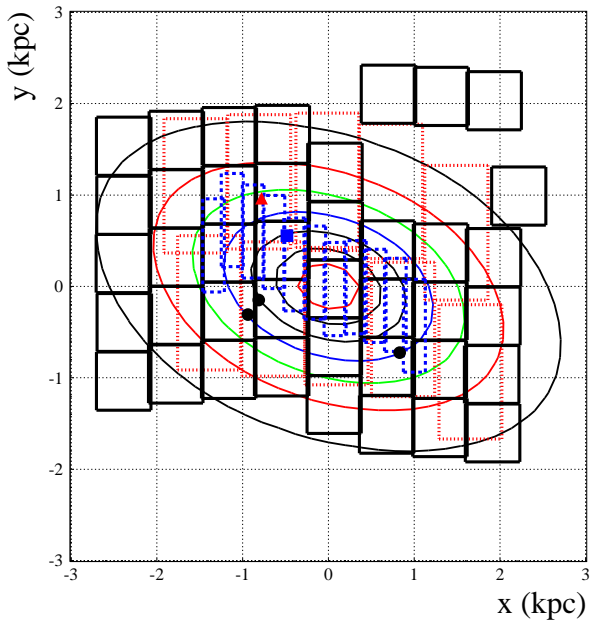}
\caption{The fields of view monitored towards the SMC
projected on the plane of the sky
by OGLE-II (dashed lines, 11 fields), OGLE-III
(solid lines, 41 fields) and EROS-2 (dotted lines, 10 fields).
The position of the five reported candidate events
is also included: one for OGLE-II (square),
three for OGLE-III (circles) and one for EROS-2 (triangle).
Further details on the events are given in Table~\ref{tab:nobs}.
Also reported, the projected density
for our fiducial SMC model (Section~\ref{sec:model}).
The contours shown correspond to the values $0.2, 0.4, 0.6, 0.8, 1.0, 1.2, 1.4$ 
in units of $10^8~\mathrm{M}_\odot~\mathrm{kpc}^{-2}$.
The $x-y$ reference system has its origin at the centre of the SMC,
the $x$-axis antiparallel to the right ascension
and the $y$-axis parallel to the declination.
}
\label{fig:campi}
\end{figure}

Microlensing observational campaigns towards the SMC have been
carried out by the MACHO, the EROS and  the OGLE collaborations.
In Fig.~\ref{fig:campi} we show the monitored
fields of view and the reported candidate events
included in the present analysis.

The EROS collaboration, with the EROS-2 set-up, observed a field of view 
covering the innermost $9~\mathrm{deg}^2$ of the SMC
during about 7 years from 1996 to 2003.
The first results of this  campaign are discussed in \cite{eros98},
with the presentation of a long-duration event, EROS2-SMC-1,
$t_\mathrm{E}\sim 120~\mathrm{d}$, which
was argued to be due, also because of the lack
of any parallax signal \citep{gould92a}
either by a large-mass object in the Galaxy halo
or by a lens lying near the source in the SMC itself.
The event optical depth was estimated to
be compatible with that expected by SMC self lensing.
This event, first reported by the MACHO collaboration \citep{macho97}
and known as MACHO 97-SMC-1, has been the object also
of a spectroscopic analysis \citep{sahu98}
whose  conclusions as for the nature of the lens, based on the lack of any signal
from the lens, excluded it from being  an MW disc star, 
are in agreement with those presented in the original
EROS analysis. A second analysis of this SMC EROS-2 campaign,
for 5 years of data, was then presented in \cite{eros03} with the inclusion
of three additional long-duration candidate events claimed however
to be \emph{doubtful}, and finally rejected in the definite
analysis presented in \cite{eros07} where
only EROS2-SMC-1 was retained as a reliable candidate event
(and with the analysis of \cite{assef06} further favouring
the SMC self-lensing interpretation of this event).
In their final analysis on the MACHO issue 
out of observations towards both Magellanic Clouds
\citep{eros07} the EROS collaboration restricted the number of sources to a subset 
of ``bright'' source objects to better address the issue of blending.
Overall, the EROS-2 SMC campaign lasted $T_\mathrm{obs}=2500~\mathrm{d}$
with an estimated total number of $0.86\times 10^6$ monitored sources. 
With no candidate events reported towards the LMC,
\cite{eros07} consider the observed rate compatible
with the expected self-lensing signal and 
get to strong constraints on the halo mass fraction in form of MACHOs.

\begin{table}
\caption{Microlensing candidate events 
for the OGLE-II, OGLE-III and EROS-2
observational campaigns towards the SMC.
The values for the duration, which are those used for the present analysis, 
and the estimate for the 
optical depth are from \protect\cite{ogle09}, \protect\cite{ogle11b} and \protect\cite{eros07},
respectively. The coordinate positions are expressed in
term of the reference frame used in Fig.~\ref{fig:campi}.
}
\begin{center}
\begin{tabular}[h]{c|c|c|c|c|c|c}
 Event & $x$ & $y$ & $t_\mathrm{E}$ & $\tau$ \\
       & [kpc] & [kpc] & d &[$10^{-7}$]\\
\hline
OGLE-SMC-01 & -0.485679 & 0.555917 & 65.0 & 1.55\\
\hline
OGLE-SMC-02 & 0.831350 & -0.725003 & 195.6 & 0.85\\
OGLE-SMC-03 & -0.937994 & -0.309247 & 45.5 & 0.30\\
OGLE-SMC-04 & -0.812418 & -0.150195 & 18.60 & 0.15\\
\hline
EROS2-SMC-1 & -0.781914 & 0.966178 & 125. & 1.7
\end{tabular}
\end{center}
\label{tab:nobs}
\end{table}

The OGLE collaboration is monitoring the SMC for microlensing
events since more than 15 years.
\cite{ogle10} reported results out of the OGLE-II campaign,
(1996-2000), covering the SMC innermost $2.4~\mathrm{deg}^2$ for a total duration 
$T_\mathrm{obs}=1408~\mathrm{d}$.
The OGLE collaboration makes
the distinction between a larger sample of ``All''
and a restricted one of ``Bright'' sources,
the latter  chosen so to reduce
the impact of blending in the analysis 
(for a discussion of the observational strategy of OGLE, in particular as for the choice
of the source sample, we refer to \citealt{novati11}).
OGLE reports an estimated number of potential sources 
$N=3.6\times 10^6$ ($N=2.1\times 10^6$), for the All (Bright) sample, respectively. 
Although  \cite{ogle10} discuss in general terms the analyses for both 
samples of sources, they specifically report the results for the All sample only.
Accordingly, this is the only one 
we will include within our analysis for OGLE-II.
In particular, \cite{ogle10} report a single-candidate event,
OGLE-SMC-01, which is considered compatible,
based on the optical depth, with the expected self-lensing signal.
Thanks to an updated set-up, a much larger SMC field of view,
$14~\mathrm{deg}^2$, was monitored during the OGLE-III phase (2001-2009).
The results of this analysis are discussed in \cite{ogle11b}.
The observational campaign lasted $T_\mathrm{obs}=2870~\mathrm{d}$
with an estimated number of sources equal
to $N=5.97\times 10^6$ ($N=1.70\times 10^6$) for the All (Bright) sample,
respectively. Three additional microlensing events are reported,
OGLE-SMC-02, OGLE-SMC-03 and OGLE-SMC-04
(with OGLE-SMC-03 belonging to the All sample only),
with the total optical depth still estimated to be in agreement
with that expected from SMC self lensing.

Among the OGLE-III SMC candidate events, OGLE-SMC-02 
(also known as OGLE-2005-SMC-1) deserved special attention. This was alerted by the
OGLE-III Early Warning System \citep{udalski03}
and enjoyed additional observations also from space, with \emph{Spitzer}, used 
to break the model degeneracies and solve the event,
with the specific aim to measure the microlensing parallax \citep{dong07}.
\cite{dong07} address in particular the issue
of the nature of the lens and conclude that the most
likely location is the Galactic halo
from a (binary\footnote{OGLE-2005-SMC-1 shows a 
deviation from that of a single lens event which
has led \cite{dong07} to conclude for a binary lens system.
The anomaly is however extremely small so that the 
event is selected in the, single lens, \cite{ogle11b} analysis.}) black hole with a total
mass of around $10~\mathrm{M}_\odot$.

For the EROS-2, OGLE-II and OGLE-III analyses
the source number is reported \emph{per field},
with 10, 11 and 41 fields monitored by each experiment,
respectively. In the following we do not include the field 140 of the OGLE-III campaign,
the isolated field in the north-west part of Fig.~\ref{fig:campi},
which presents a strong over-density of (potential lens) stars being
centred along the line of sight of the foreground 47~Tuc (NGC104) globular cluster.

\begin{figure}
\includegraphics[width=84mm]{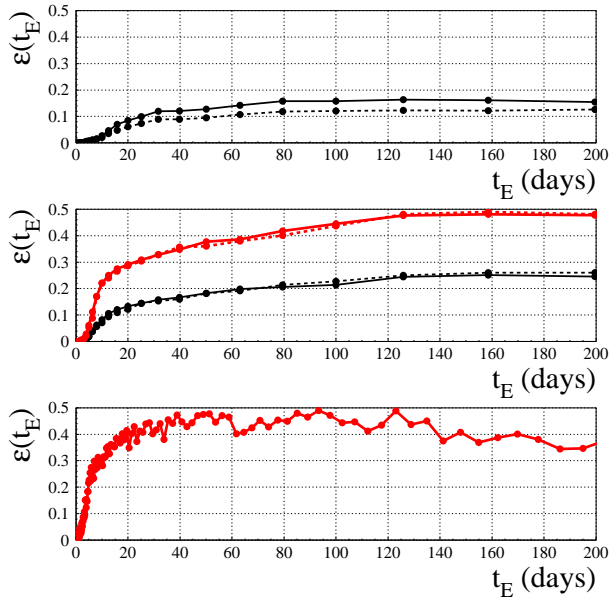}
\caption{The detection efficiency 
as a function of the duration, ${\cal{E}}(t_\mathrm{E})$,
for OGLE-II (top panel, All sample), OGLE-III (middle panel) and EROS-2.
For OGLE the solid and dashed curves trace 
the efficiency for ``sparse'' and ``dense'' fields,
as a measure of the crowding, respectively
(for OGLE-III the two curves are almost indistinguishable).
For OGLE-III the thicker curves (with larger values 
of the efficiency) refer to the Bright sample of sources.
}
\label{fig:eff}
\end{figure}
Both  EROS \citep{eros07} and OGLE \citep{ogle10,ogle11b}
carry out an analysis of their detection efficiency
based on the estimated number of monitored sources
and presented in terms of the event duration, ${\cal{E}} = {\cal{E}}(t_\mathrm{E})$,
which we also include in our analysis. 
In particular, OGLE reports the estimate for the efficiency
both for a ``sparse'' and a ``dense'' field, depending
on the density of stars. Accordingly, for each
given value of the duration, we linearly interpolate 
the efficiency taking into account the estimated number
of sources per field (the same value we use
to estimate the expected number of events), while keeping the reported values
as fixed for those fields with lower, respectively higher, source star number.
(This expedient, however, is effective
for OGLE-II only, as for OGLE-III the available
data for the sparse and dense fields
indicate that the efficiency is roughly constant
across the overall monitored field of view,
even though the choice of the, two nearby, fields
used for this analysis by OGLE-III
may have biased this outcome.)
In Fig.~\ref{fig:eff} we report 
a detail  of the efficiency function ${\cal{E}}(t_\mathrm{E})$
for $t_\mathrm{E}<200~\mathrm{d}$.
Besides the dependence on the crowding, we remark the low maximum value 
of ${\cal{E}}(t_\mathrm{E})$ for OGLE-II as compared to those of OGLE-III
and EROS-2 and, for EROS-2, the much faster increase up to rather large values 
for small durations as compared to OGLE.

Overall, there are five microlensing candidate events 
reported towards the SMC upon which EROS and OGLE
based their analyses and whose characteristics we summarize
in Table \ref{tab:nobs} 
and which we will further consider in the present analysis.
For definiteness, we will consider as homogeneous
the All sample of sources of OGLE-II and OGLE-III
and the Bright sample of sources of OGLE-III
together with that of EROS-2.

Besides EROS and OGLE, also the MACHO collaboration monitored the SMC
for microlensing events. The microlensing event MACHO Alert 98-SMC-1
has been the first binary caustic crossing event reported
towards the Magellanic Clouds \citep{macho99}, also monitored
by the PLANET collaboration \citep{rhie99}. The analysis
of the event, including additional data from the EROS and the OGLE data base, 
and in particular of the lens projected velocity,
led to the conclusion that the event is more likely to reside
in the SMC than in the Galactic halo \citep{macho99,albrow99,rhie99}.
The MACHO collaboration, however, did not present a detailed and complete analysis of the
SMC campaign, as they did for the LMC one. In particular, the estimate of the number of sources
and the analysis of the detection efficiency,
essential information to reliably assess the characteristics of the expected signal,
are both missing. For this reason hereafter we no longer consider the results of 
the MACHO collaboration campaign towards the SMC.

\section{Analysis} \label{sec:analysis}

\subsection{The microlensing optical depth} \label{sec:tau}

The optical depth, $\tau$, is the instantaneous probability to observe
a microlensing event. This is calculated as the integrated number
of potential lenses within the microlensing tube for a given
line of sight (for the background theory
of microlensing we refer to \cite{roulet97,mao12} and references therein). 
The cross-section radius of the microlensing tube is the \emph{Einstein radius}
\begin{equation}\label{eq:re}
R_\mathrm{E}=\sqrt{\frac{4Gm}{c^2}\,\frac{D_l(D_s-D_l)}{D_l}}\,,
\end{equation}
where $m$ is the lens mass and $D_l$ ($D_s$) the lens (source)
distance from the observer, respectively. 
A relevant outcome of the microlensing theory
is that the optical depth turns out to be independent from the lens mass
(for a fixed overall mass of a lens population,
lenses of smaller mass are more numerous
but have a smaller cross-section,
whereas heavier lenses are less numerous but with
a larger cross-section, and this just at the level
that the two effects compensate one each other).
Further taking into account the source density
distribution (which is relevant, in our case, for lenses within the SMC)
\begin{equation}\label{eq:tau}
\tau=\frac{4\pi G}{c^2}\frac{\int\mathrm{d}D_s\int^{D_s}\mathrm{d}D_l\rho_s \rho_l 
\frac{D_l (D_s-D_l)}{D_s}}{\int\mathrm{d}D_s\rho_s}\,,
\end{equation}
where $\rho_l$ ($\rho_s$) are the lens (source) mass density distribution, 
respectively. Hidden within the integrands of Eq.~\ref{eq:tau} there is a term $D_s^{2-2\beta}$,
both in the numerator and in the normalization in the denominator,
introduced to properly take into account  the variation with the distance of the number
of available source stars, with our choice being for the value $\beta=1$.
The reason for this term is twofold:
besides the variation of the volume element with the distance, 
which gives $D_s^2$, the term in $\beta$ follows
under the assumption that the fraction of stars
brighter than a given luminosity $L$, we recall that
microlensing surveys are magnitude-limited, 
is proportional to $L^{-\beta}$ \citep{kiraga94}.
\cite{zhao95} estimate $\beta=1\pm 0.5$, with smaller values, 
found moving towards fainter magnitude limits,
making larger optical depth and viceversa, and
$\beta=1$ being thereafter the  choice of reference 
in particular for studies towards the LMC/SMC 
(we also refer to the discussion of this issue
in \cite{gyuk00} specific to the case of the Magellanic Cloud).
Indeed, considering the luminosity function reported
by OGLE towards the SMC \citep{ogle10}, we find agreement
to this value within 10\%. For reference, we evaluate  
variations of the SMC self-lensing optical depth
below 1\% and of about 4\% varying $\beta$
of $10\%$ and $50\%$, respectively, from $\beta=1$.
Note that this is not in disagreement 
with the significant variations of $\tau$ with respect
to $\beta$ reported towards the Galactic Bulge \citep{kiraga94,hangould03}.
Indeed, the extent of the dependence on $\beta$
decreases with the distance of the sources,
being already very small at the SMC distance.

According to its definition as an instantaneous
probability, $\tau$ is a static quantity
which can not be used to characterize the observed events.
This feature makes the optical
depth a very useful quantity from a theoretical point of view,
being less model dependent,
but also observationally. An estimate
of the measured optical depth can indeed
be used to trace the underlying mass (and spatial) density
distribution of a given lens population.

For an experiment with overall
duration $T_\mathrm{obs}$ and $N_\mathrm{obs}$ observed sources 
and sensitive to event up to maximum magnification
$u_0(\mathrm{max})$ ($u$ being the impact parameter,
the distance of the line of sight to the
lens trajectory, which is
roughly inversely proportional to the magnification at maximum), 
for a set of $N_\mathrm{ev}$ observed events
with duration\footnote{The timescale
of a microlensing event is the \emph{Einstein time},
$t_\mathrm{E}=R_\mathrm{E}/v$ where $v$ is
the (transverse component of the) 
lens velocity with respect to the microlensing tube.}
$t_{\mathrm{E},i}$ (with $i=1,\ldots,N_\mathrm{ev}$)
and given efficiency ${\cal E}(t_{\mathrm{E},i})$, the estimate
of the \emph{measured} optical depth reads
\begin{equation}\label{eq:tau_obs}
\tau_\mathrm{obs} = \frac{\pi}{2\,u_0(\mathrm{max}) N_\mathrm{obs} T_\mathrm{obs}}\,
\sum_i^{N_\mathrm{ev}}\frac{t_{\mathrm{E},i}}{{\cal E}(t_{\mathrm{E},i})}\,,
\end{equation}
with the associated statistical error evaluated through the
prescription of \cite{hangould95b} 
\begin{equation}\label{eq:tau_sigma}
\sigma(\tau) = \tau\,\frac{\sqrt{<t^2_\mathrm{E}/{\cal E}^2>}}{<t_\mathrm{E}/{\cal E}>}\,
\frac{1}{\sqrt{N_\mathrm{ev}}}\,.
\end{equation}

\begin{figure}
\includegraphics[width=84mm]{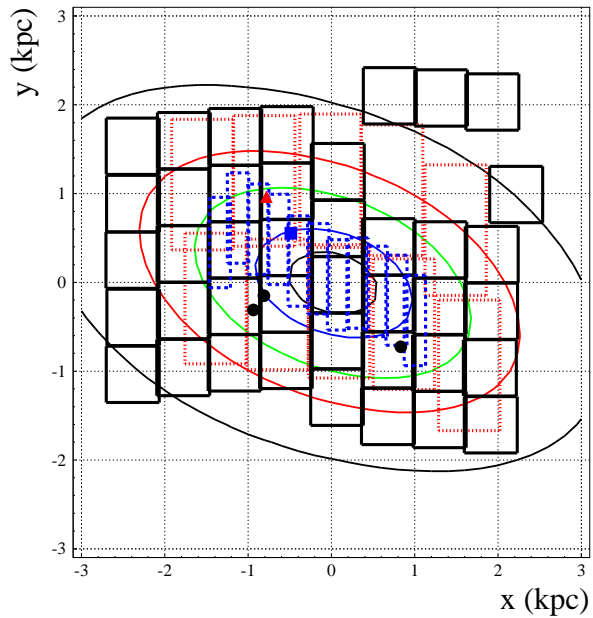}
\caption{SMC self-lensing optical depth profile.
The contours shown correspond to the values
$0.1,0.3,0.5,0.8,1.0$ in units of $10^{-7}$.
The maximum value is $1.3\times 10^{-7}$.
The reference system, the observed event positions and field
contours are indicated as in Fig.~\ref{fig:campi}.
}
\label{fig:tau_sl}
\end{figure}

Coming to the specific problem of SMC microlensing,
also looking at Eq.~\ref{eq:tau},
we expect the signal from the MW
lens populations to be rather independent from
the inner structure of the SMC (with $\rho_l\approx 0$ 
within the SMC where $\rho_s \ne 0$). 
It results\footnote{For cross-check, the evaluation of the optical depth profiles
has been carried out independently by two of us.}
, in particular, that the profiles for the MW disc and the Galactic halo
optical depth are roughly constant across the field of view.
Specifically, for the MW halo profile $\tau\sim 6.3\times 10^{-7}$
(for a full MACHO halo)
and for the MW disc $\tau\sim 0.04\times 10^{-7}$,
in both cases with relative variations up to 5\% level.
The SMC self-lensing optical depth, on the other hand,
following the underlying lens spatial density profile,
is strongly variable, Fig.~\ref{fig:tau_sl},
with peak central value, for our fiducial model,
$\tau=1.3\times 10^{-7}$, and the observed events
falling within the lines of equal optical depth values
$0.5$ and $0.8\times 10^{-7}$. 
As expected, the introduction of a shift in distance between
the OS and the YS population for a test model
against the fiducial one (Section~\ref{sec:structure})
enhances the SMC self-lensing signal. The relative
increase with respect to the fiducial model
is at 6\% level at the SMC centre
and below 5\% for the average values across
the monitored fields of view
(in particular, with the YS lying $2~\mathrm{kpc}$
behind the OS, there is a strong enhancement, about 80\%,
of the signal from YS sources with OS lenses,
which is however almost completely compensated
by a corresponding decrease in the signal
from OS sources with YS lenses).
For completeness we mention  also the outcome of the optical depth
analysis for the SMC dark matter halo.
The profile is asymmetric following the underlying
SMC luminous profile and overall inclination,
with, for a full SMC halo, peak value $0.46\times 10^{-7}$
(in the south-west part of the SMC, following the
SMC inclination, around at position $x,y=1.1,-0.6$
in the reference frame of Figs.~\ref{fig:campi} and \ref{fig:tau_sl})
and average value across the field of view 
in the range $(0.31-0.38)\times 10^{-7}$ (the smaller
and larger value for OGLE-III and OGLE-II fields, respectively).
Overall, this is only about 5\% of the MW halo signal
and therefore we will hereafter neglect this component.

The MW dark matter halo optical depth we evaluate
towards the SMC for a full MACHO halo, $6.3\times 10^{-7}$,
is significantly larger than the corresponding value
we had evaluated towards the LMC, $4.5\times 10^{-7}$ \citep{novati09b}.
This increase is to be attributed
to the increase in Galactic longitude and,
to somewhat less extent, to the increase of the distance
(whereas the increase, in absolute value, in Galactic latitude 
tends to reduce the optical depth).

In the following we will address the issue of the nature
of the observed events through the analysis
of the microlensing rate. It is however useful
to consider, to some extent, this issue already within the framework of the optical
depth, in particular asking whether the stellar lens populations
may or may not explain the observed signal and
this starting from the consideration that the
largest signal is expected, as it may be guessed
looking at the relative values of the optical
depth, from stellar lenses within the SMC rather than
from MW disc lenses. 

The expected quantity to be compared
to the measured optical depth is the \emph{average}
optical depth value across the field
of view where the $N_\mathrm{obs}$ source stars
entering Eq.~\ref{eq:tau_obs} are monitored. 
Furthermore, as to be expected and according
to Eq.~\ref{eq:tau_sigma}, the relative (statistical) error
on this estimate scales with the (square root of the) number of 
observed events and in particular 
$\sigma(\tau)/\tau = 1$ for $N_\mathrm{ev}=1$.
To draw robust conclusions based on the optical depth,
being statistical statements,
a large enough sample of observed events is therefore mandatory.

The average optical depth for SMC self lensing
across the monitored fields of view, according to our fiducial
model, is $<~\tau_\mathrm{exp}~> = 0.50$, $0.81$ and $0.39$ (in units of $10^{-7}$)
for EROS-2, OGLE-II and OGLE-III, 
respectively. This quantity is to be compared with the values
already reported in Section~\ref{sec:obs}.
For EROS-2 and OGLE-II, with a unique 
event (and always in units of $10^{-7}$) $\tau_\mathrm{obs}=1.7\pm 1.7$ (EROS-2)
and $1.55\pm 1.55$ (OGLE-II) and  $\tau_\mathrm{obs}=1.30\pm 1.01$
for OGLE-III with three reported events.
Although larger, the observed values are, within their large error,
easily in agreement with the expected ones for SMC self lensing.

We can compare these results with those reported 
in the analysis of \cite{eros98}, as for the SMC self lensing
signal, often quoted and used as a ``fiducial value''.
In particular, for a scale length $h=2.5~\mathrm{kpc}$ 
\cite{eros98} report, for SMC self lensing, an average value of $1.0\times 10^{-7}$.
Allowing for the caveat of the different
central normalization (Section~\ref{sec:smc}),
possibly because of a different definition
of the region over which we average the optical depth
and/or for a difference in other parameters of the model,
we fail to reproduce this result. With a peak central value
of $6.2\times 10^{-7}$ we find instead
an average expected value of $0.8\times 10^{-7}$ for the EROS-2
monitored fields of view. For OGLE-II and OGLE-III
we obtain $1.7$ and $0.56\times 10^{-7}$, respectively.
Comparing with the results of our fiducial model,
following also the discussion in Section~\ref{sec:smc},
we find that this density distribution leads to a much more
centrally peaked optical depth profile
(with the larger relative difference for the OGLE-II fields).

The SMC self-lensing optical depth has been
analysed also by other authors.
\cite{sahu98} estimate values in the range $1.0-5.0\times 10^{-7}$.
\cite{graff99}, based on the \cite{gardiner96} $N-$body simulation
of the SMC, derived an average smaller value,
$0.4\times 10^{-7}$, arguing that, compared
to the \cite{eros98} and \cite{sahu98}, both reporting
larger values, a reason of disagreement could be traced back
in the smaller line-of-sight thickness used.
All these analyses, however, somehow suffer
from the very large uncertainties in the model 
of the SMC luminous components which,
if not still fully solved, are by now
largely smoothed out by the more recent analyses
(as in particular those based on the newly available OGLE-III data set).

\subsubsection{The SMC self-lensing optical depth: a spatial
distribution analysis} \label{sec:tau_bin}

Given the very small number of observed events
we can not aim at drawing
stronger conclusions on the basis of 
the optical depth analysis. We can still,
however, try to gain some further insight
by addressing the issue of the spatial
distribution of the events. The motivation comes
from the observation of the very rapid,
non-linear, variation of  the expected
optical depth profile across the monitored fields of view
(which is made apparent, for instance, by the strong decrease of the expected
average value moving from OGLE-II to OGLE-III, where a much
larger region has been monitored). This makes the
average optical depth value reported above of limited interest.
The usual way out to address this issue
is to consider smaller and more homogeneous sub-regions
where to perform the analysis.

Our choice is to fix the bin size by asking that each bin contain
the  same number of monitored source stars (this is
suggested  by Eq.~\ref{eq:tau_obs} according to which,
once fixed the observed events, the estimate
of the observed optical depth scales as $1/N_\mathrm{obs}$).
Specifically, we choose to select four bins
whose exact extension varies according to
the different set-up we consider (EROS-2, OGLE-II and OGLE-III).
The values of the average expected optical
depth for SMC self lensing within the bins
are reported in Table~\ref{tab:tau_bin}
(for convenience all the values are divided by a factor 4
so that they can be directly compared to
the already reported observed values,
as each bin contains exactly $1/4$ of the 
total number of monitored sources).
The data in Table~\ref{tab:tau_bin}
allows us to better quantify 
the variation of the optical depth
across the monitored field of views shown in 
Fig.~\ref{fig:tau_sl}. In particular
we note the much stronger gradient
expected with the \cite{eros98} model
as compared to the smoother fiducial one.
Additionally, we can trace the position
of the observed events within this parameter space,
providing us with a more quantitative hint
on their spatial distribution
(a larger set of events would then allow one
to carry out a more robust analysis
by comparing the observed to the expected 
profile of the optical depth).
In particular, it results that, moving
from the outer bin inwards,
both the EROS-2 and OGLE-II events
fall within the second bin
(which, especially for the second event,
at glance from Fig.~\ref{fig:tau_sl} is not apparent)
whereas all the three OGLE-III events
fall within the third bin (the corresponding
values are underlined in Table~\ref{tab:tau_bin}), therefore
in a more central position.

\begin{table}
\caption{
Expected average values for the SMC self-lensing
optical depth within the bins
of the spatial distribution analysis.
In columns (1-3) and (4-6) we report the results
for the fiducial and the \protect\cite{eros98} models,
respectively. In particular we report
the results for EROS-2, columns (1) and (4); OGLE-II, columns (2) and (5); 
OGLE-III, columns (4) and (6).
The observed event(s) for each data set fall within the bin
whose expected value is underlined.
The reported values are normalized so that they
are homogeneous to the observed values reported
in Table~\ref{tab:nobs} (see the text for further details).
}
\begin{center}
\begin{tabular}[h]{c|c|c|c|c|c|c}
Bin & (1) & (2) & (3) & (4) & (5) & (6)\\
\hline
1 & 0.08 & 0.17 & 0.06 & 0.06 & 0.19 & 0.03\\
2 & \underline{0.13} & \underline{0.21} & 0.11 & \underline{0.15} & \underline{0.43} & 0.12\\
3 & 0.17 & 0.24 & \underline{0.17} & 0.31 & 0.65 & \underline{0.27}\\
4 & 0.22 & 0.28 & 0.24 & 0.55 & 0.96 & 0.65\\
\end{tabular}
\end{center}
\label{tab:tau_bin}
\end{table}

For the above discussion on the spatial distribution
we have considered, for self lensing,
the SMC luminous component lenses only. 
This is justified by the much larger
optical depth of this component compared
to that of MW disc lenses. Specifically,
the ratio of the optical depth average value for SMC self lensing over
that of MW disc vary in the range from $\sim 10$ up to 
$\sim 20$ (for OGLE-III and OGLE-II fields, respectively,
the second being more clustered around the SMC centre).
As further addressed below, coming to the 
expected signal in term of number of events,
the SMC self-lensing signal remains larger
than that of the MW disc lenses, but only about half as large
as it results from the optical depth analysis alone.

\subsection{The microlensing rate} \label{sec:rate}

The microlensing rate, $\Gamma$, is defined as the number of new lenses, per unit time,
entering the microlensing tube and therefore giving rise to a new microlensing event,
for a given line of sight and per source star. It is therefore a \emph{dynamic} quantity, as opposed
to the optical depth. We recall that for a generic microlensing event,
point-like single lens and source with uniform relative motion,
the only physically available measured parameter of the lensing
parameter space, besides the position, is the event duration, $t_\mathrm{E}$.
In particular, the lens mass, the lens and source distances
and the relative velocity are not directly accessible to the observations.
At the price of introducing a number of additional
ingredients in the model, with respect to the optical depth,
the microlensing rate provides us with the expected
event number and, in particular,  the expected 
duration and position distributions,
\begin{eqnarray} \label{eq:rate}
\mathrm{d}\Gamma &=& 2 \rho_l(D_l) \frac{\rho_s(D_s)}{\mathrm{I}_s} u_0(\mathrm{max})\times\nonumber\\
&&\xi(\mu) v R_\mathrm{E}(D_l,D_s,\mu) P(v)
\mathrm{d}D_l\mathrm{d}D_s \mathrm{d}\mu \mathrm{d}v\,,
\end{eqnarray}
where $\mathrm{I}_s$ is the normalization for the source density distribution,
the integration of $\rho_s$ along the line of sight, $\xi(\mu)$ the lens mass
function. $P(v)$ is the (assumed isotropic) distribution for the lens-source relative velocity
(transverse to the line of sight)
\begin{equation} \label{eq:pv}
P(v) = \frac{1}{\sigma^2_{sl}}\,v \exp\left(-\frac{v^2+A^2}{2\sigma^2_{sl}}\right)\,
\mathrm{I}_0\left(\frac{A v}{\sigma^2_{sl}}\right)\,,
\end{equation}
where $\mathrm{I}_0$ is the modified Bessel function of first kind,
$\sigma^2_{sl} \equiv \sigma_l^2+x^2 \sigma_s^2$, with $\sigma_l$ ($\sigma_s$) the
lens (source) 1-d velocity dispersion, $A$ the modulus of the bulk
motion components (solar motion, SMC internal and bulk motions), $x\equiv D_l/D_s$
(for a discussion we refer for instance to \cite{novati08} where also
the more general case of an anisotropic Gaussian distribution is addressed).

The number of expected events, $N_\mathrm{exp}$, is proportional to
the integral of the microlensing rate over the full
available parameter space. The experimental detection efficiency being
usually evaluated as a function of the event duration, ${\cal{E}}={\cal{E}}(t_\mathrm{E})$,
it results
\begin{equation} \label{eq:nexp}
N_\mathrm{exp} = N^*_\mathrm{obs} T_\mathrm{obs} 
\int{\mathrm{d}t_\mathrm{E} \,\frac{\mathrm{d}\Gamma}{\mathrm{d}t_\mathrm{E}}\,
{\cal{E}}(t_\mathrm{E})}\,.
\end{equation}
The product $N^*_\mathrm{obs} T_\mathrm{obs}$ is sometimes
referred to as the ``exposure'', $E$. 
Starting from the relation $t_\mathrm{E}=R_\mathrm{E}/v$
we evaluate, from Eq.~\ref{eq:rate}, 
$\mathrm{d}\Gamma/\mathrm{d}t_\mathrm{E} = 
\mathrm{d}\Gamma/\mathrm{d}v \times R_\mathrm{E}/t_\mathrm{E}^2$.

\subsection{The number and the duration of the expected events} \label{sec:res}

In this section, we establish the basis for our following analysis
on the lens nature for the observed events by reporting
the results we obtain by the analysis of the microlensing rate.

As a first step we evaluate the differential rate 
$\mathrm{d}\Gamma/\mathrm{d}t_\mathrm{E}$ 
for all the populations we consider: SMC self lensing, MW disc and
MACHO lenses. The number of sources, for each experiment, being known \emph{per field}, 
we therefore evaluate the rate towards the 
central line of sight of each EROS-2, 
OGLE-II and OGLE-III field. For SMC self lensing, because of the large variation
across the monitored fields of the underlying lens population,
we rather evaluate the \emph{average} rate across the field of view. 
This becomes relevant especially
for the more peaked \cite{eros98} model,
with an overall decrease in the number
of expected events, relative to the 
case where the single central line of sight is considered,
that sums up to about 10\% (and is much larger in the innermost fields).

\begin{table}
\caption{Microlensing rate analysis: 
expected duration distribution
for self lensing lenses and MW MACHO lensing.
We report the  $16\%, 34\%, 50\%, 68\%, 84\%$ values
for the OGLE-III All sample
set-up and detection efficiency.
}
\begin{center}
\begin{tabular}[h]{cccccc}
lenses& 16\% & 34\% & 50\% & 68\% & 84\%\\
& (d) & (d) & (d) & (d) & (d) \\ 
\hline
SMC          &  46. &  65. & 84. & 110. & 150. \\
SMC BD       &  19. &  25. & 31. &  40. &  54.  \\
MW disc      &  25. &  36. & 48. &  66. &  94.  \\
MW disc BD   &  11. &  15. & 19. &  24. &  33.  \\
SL           &  34. &  53. & 71. &  98. & 140. \\
\hline
$10^{-3}~\mathrm{M}_\odot$ &  2.9 &  4.2 &  5.4 &   7.2 & 10.\\
$10^{-2}~\mathrm{M}_\odot$ &  5.7 &  7.4 &  9.0 &  12.  & 16.\\
$10^{-1}~\mathrm{M}_\odot$ & 12.  & 16.  & 20.  &  27.  & 37.\\
$1~\mathrm{M}_\odot$       & 34.  & 48   & 60.  &  78.  & 110.\\
$10~\mathrm{M}_\odot$      & 100. & 140. & 170. & 220.  & 300.\\
\hline
\end{tabular}
\end{center}
\label{tab:te}
\end{table}
In Table~\ref{tab:te} we report some statistics
on the expected duration distribution
for self lensing populations and MACHO
lensing for the OGLE-III All sample set-up
and detection efficiency.
As remarked, the EROS-2 corresponding distribution
is somewhat shifted towards
smaller values of $t_\mathrm{E}$,
at about 10\%-20\% level.

\begin{figure}
\includegraphics[width=84mm]{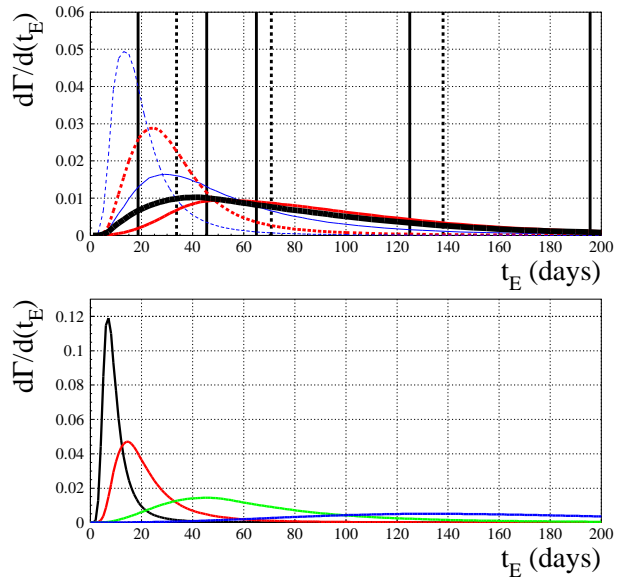}
\caption{Normalized differential rate distribution,
$\mathrm{d}\Gamma/\mathrm{d}t_\mathrm{E}$,
corrected for the detection efficiency.
Top panel: the expected distribution, each separately normalized,
for the different self-lensing populations considered. 
Dashed and solid line are for the brown
dwarf and star lenses,
thin and thick lines for MW disc and SMC lenses.
The thicker solid line is for the resulting overall
self-lensing distribution.
The dotted vertical lines indicate
the 16\%, median and 84\% values of this distribution.
The solid vertical lines indicate
the values for the observed events, Table~\ref{tab:nobs}.
Bottom panel: the expected
distribution for MW MACHO lenses
varying the MACHO mass.
Moving from left to right as for the modal value:
$0.01,\,0.1,\,0.5$ and $1~\mathrm{M}_\odot$.
}
\label{fig:dgte}
\end{figure}

In Fig.~\ref{fig:dgte} we show the differential rate
modulated by the detection efficiency,
$({\mathrm{d}\Gamma}/{\mathrm{d}t_\mathrm{E}})_{\cal E}$
for (both stars and brown dwarfs) SMC self lensing and MW disc lenses 
(top panel) and for a set of MACHO mass values
from $10^{-2}$ up to $1~\mathrm{M}_\odot$
for MW MACHO lenses.
The (normalized) distributions  shown are averaged
across the monitored fields of view
(the spatial variation being more pronounced 
for the SMC self-lensing signal).
In particular we show the result we obtain
in the OGLE-III case.

In Table~\ref{tab:nevt} we report the total number of expected events
for the three experiment considered, EROS-2, OGLE-II and OGLE-III,
for the self-lensing population considered, SMC self lensing
and MW disc lenses, for both the stellar and brown dwarf
contribution, for the All and Bright sample, whenever the case.
\begin{table}
\caption{Microlensing rate analysis: expected number 
of events for the self lensing populations 
(BD stands for brown dwarfs)
for each of the three experiment analysed.
}
\begin{center}
\begin{tabular}[h]{ccccc}
lenses&OGLE-II&\multicolumn{2}{c}{OGLE-III}&EROS-2\\
& All & All & Bright & Bright\\
\hline
SMC & 0.36 & 1.25 & 0.71 & 0.33\\
SMC DB & 0.036 & 0.13 & 0.079 & 0.045\\
MW disc & 0.045 & 0.22 & 0.13 & 0.077 \\
MW disc BD & 0.0035 & 0.020 & 0.014 & 0.0093\\
\hline
& 0.44& 1.62 & 0.93 & 0.46\\
\end{tabular}
\end{center}
\label{tab:nevt}
\end{table}
The inspection of this table suggests a few comments.
As for the relative weight of the different experiment,
for the All sample of sources, the OGLE-III
expected signal is about three times larger
than that of OGLE-II. The EROS-2 signal
sums up to about half of that of the
Bright sample of OGLE-III. The MW disc
signal is overall rather small compared
to the SMC self-lensing one.
The different relative weight for OGLE-II
and OGLE-III (about 7\% against 12\%)
can be traced back mainly to the different
extent of the monitored fields. The somewhat larger
fraction for EROS-2, 17\%, can be understood on
the basis of the larger efficiency for
smaller values of the Einstein time.
Overall, this makes EROS-2 quite relevant
as compared to OGLE-III. Finally, although small, 
the expected SMC brown dwarf signal
turns out to be about as large of the stellar
MW disc one. Here again the relative increase
for EROS-2 can be traced back to the different
shape of the efficiency curve.
Overall, the MW disc signal represents
10\%-16\% of the overall self-lensing signal.
The enhancement of this ratio when considering
the expected number as compared to the
optical depth analysis is understood,
given that $\Gamma \propto \tau/t_\mathrm{E}$,
on the basis of the expected shorter 
duration of MW disc events.

The \cite{eros98} model strongly
enhances the SMC self-lensing
expected signal, resulting
in about twice as much expected events.
Coherently with the optical depth analysis 
these are found to be, however, much more
strongly peaked  in the innermost
SMC region (for OGLE-III, for instance,
we find that 60\% of the events should
be expected in the innermost bin,
defined as in the previous optical depth analysis,
against 40\% for our fiducial model).
In particular, the expected number
of self-lensing events is 1.0 (OGLE-II),
3.1 and 1.8 (OGLE-III, All and Bright sample, respectively)
and 0.8 (EROS-2). The major  enhancement (about 2.5 times
as much) is found, as expected, for the more
centrally clustered OGLE-II fields of view.

\begin{figure}
\includegraphics[width=84mm]{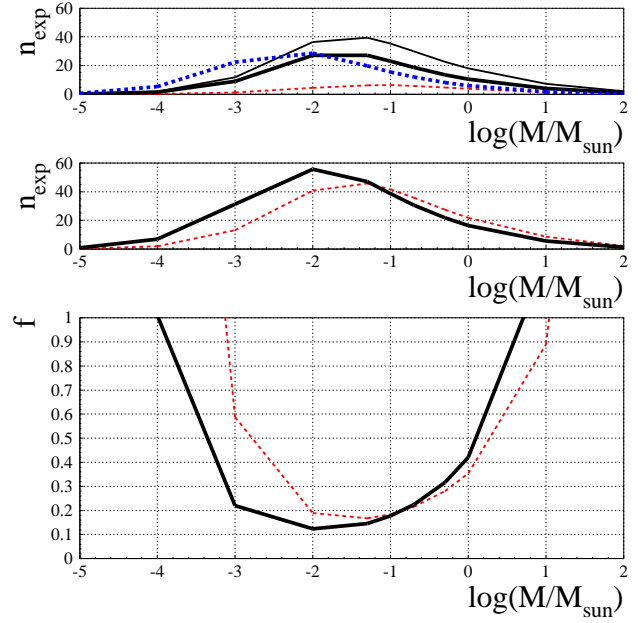}
\caption{Top and middle panel: 
number of expected MW MACHO lenses
events as a function of the MACHO mass
for a full MACHO halo.
Top panel: we report separately the results 
for OGLE-II (All sample), OGLE-III (All and Bright
samples) and EROS-2 (dashed, thin solid, thick
solid and dot-dashed lines, respectively).
Middle panel: we report separately
the results for the All sample (OGLE-II and OGLE-III,
dashed line) and the Bright sample
(OGLE-III and EROS-2, solid line).
Bottom panel: 95\% CL upper limit
for the halo mass fraction
in form of MACHOs based on the Poisson
statistics of the number of events
(see the text for details).
Solid and dashed curves as in the middle panel.
}
\label{fig:nexp}
\end{figure}

The number of expected dark matter events, as a function of the MACHO mass
for a full MACHO halo, is shown in Fig.~\ref{fig:nexp}.
The expected increase in the number for smaller values of the MACHO mass 
is increasingly compensated by the corresponding
decrease in the detection efficiency for small values of the event duration. 
In particular, coherently with the relative difference
in their detection efficiency functions ${\cal E}(t_{\mathrm{E}})$, 
the expected EROS-2 signal overtakes the OGLE-III one for
values below $5\times 10^{-3}\mathrm{M}_\odot$.
Overall, the expected signal rapidly drops to zero
below $10^{-3}-10^{-4}\mathrm{M}_\odot$ and,
at the opposite end, above $1-10~\mathrm{M}_\odot$.
The expected peak values in the number of events is for a MACHO population
in the mass range  $10^{-2}-10^{-1}\mathrm{M}_\odot$. 

Overall, the relative ratios of the expected number of events from
OGLE-II, OGLE-III (All and Bright sample) and EROS-2 are well understood
starting from Eq.~\ref{eq:nexp} and the specifications
of the different set-up, in particular the value
of the exposure, $E$ and the efficiency curve ${\cal E}(t_\mathrm{E})$.
OGLE-II enjoys a large exposure, $E_\mathrm{OGLE-II}=5.1\times 10^9$, however
it suffers from a quite small efficiency, below 10\% for
$t_\mathrm{E}<10~\mathrm{d}$ and rising at most up to about
16\%. For OGLE-III it results $E_\mathrm{OGLE-III}=1.7\times 10^{10}$,
for the All sample
and $4.9\times 10^9$ for the Bright sample
for which, on the other hand, the efficiency is up to about
twice as large than for the All sample. In particular, 
${\cal E}(t_\mathrm{E})\sim 20\%~(10\%)$ for
$t_\mathrm{E}=10~\mathrm{d}$, $\sim 30\%~(13\%)$
at $20~\mathrm{d}$ with top values about
$50\%~(25\%)$ in the range $\sim 120-300 ~\mathrm{d}$,
for the Bright (All) sample, respectively.
Finally, EROS-2 is characterized by a long duration
but a relatively small number of monitored
sources so that $E_\mathrm{EROS-2}=2.3\times 10^9$,
half as small $E_\mathrm{OGLE-III}$ Bright sample.
The strength of EROS-2 is however the efficiency
reaching 30\% already at $t_\mathrm{E}\sim 10~\mathrm{d}$
and remaining stable above 40\% in the range $t_\mathrm{E}\sim 20-140~\mathrm{d}$.
On this basis we can understand, for instance,
the large number of EROS-2 expected MACHO lensing events, in particular
for low mass values ($10^{-3}-5\times 10^{-2}\,\mathrm{M}_\odot$),
as compared to OGLE-II whereas the expected 
self-lensing signal, for EROS-2 and OGLE-II, turns out
to be completely equivalent in term of the number of expected events.

In the bottom panel of Fig.~\ref{fig:nexp}, we report
the 95\% CL upper limit for the halo mass fraction
in form of MACHOs, $f$, based on the Poisson
statistics of the expected versus the observed number of events.
In particular, we make use
of the confidence level statistics for
a Poisson distribution with  a background,
also following a Poisson distribution, 
whose mean value is supposed to be exactly
known and which is given in our case by the expected 
self-lensing signal, following the recipe
of \cite{FC98}. This gives us, in particular,
the upper limit, fixed the confidence level,
for the signal (the MACHO lensing number of events).
We consider separately the full set of
the All sample of sources (OGLE-II and OGLE-III)
and the Bright one (OGLE-III and EROS-2).
For the All (Bright) sample with $n_\mathrm{obs}=4$ $(3)$ reported
candidate events and a background signal
of $n_\mathrm{exp,SL}=2.06$ $(1.39)$ the 95\% CL
upper limit turns out to be of 7.70 (6.86) events.
The lowest upper limit (here and in the following
at 95\% CL) for the Bright (All) sample is for $10^{-2}$
($5\times 10^{-2}$) $\mathrm{M}_\odot$
at $f=12\%$ ($17\%$), with $f=32\%$ ($28\%$)
for $0.5~\mathrm{M}_\odot$, respectively.
The profile of the upper limit
for the All and Bright sample follow,
reversed, that of the expected number of MACHO lensing events
modulated by the expected background signal values.
In particular, when joining OGLE-II and OGLE-III for the All
sample and OGLE-III and EROS-2 for the Bright sample,
following the already remarked enhanced
efficiency of EROS-2 to short duration (low-mass)
events, the resulting constraints for $f$ 
are stronger (also in an absolute sense) for the Bright sample for
small mass values (here roughly below $0.1~\mathrm{M}_\odot$),
respectively stronger for the All sample above this threshold.

In their analyses, the OGLE collaboration,
roughly based on the expected optical depth
but lacking an explicit evaluation of the expected
number of event for the self-lensing signal, 
and also following \cite{moniez10},
assumes that the background (self lensing) expected 
value is equal to the number of observed events.
In this case four (three) for the All (Bright) sample,
against our values, 2.1 (1.4), respectively.
Under this assumption the upper limits for the signal,
and therefore those on $f$,
are accordingly smaller, in this case 5.76 (5.25), respectively
(which makes, in relative terms, a rather significant change).
For reference we mention the values of these same upper limits
assuming, instead, that the expected background is zero
(namely, assuming that there is no expected self lensing signal),
9.76 and 8.25 for the All and Bright samples, respectively.

Starting from the larger values of expected
self-lensing events with the \cite{eros98} model,
4.17 (2.59) for the All (Bright) sample,
we would get to considerably smaller
upper limit for the Poisson statistics
with a background, 5.60 and 5.66 for the
All and Bright sample, respectively
(here the statistics makes the first value
smaller, which is opposite to the result
we obtain with our fiducial model).
This then gives rise (the expected number of MACHO lensing
events does not change) to stronger constraints
for $f$ (always at 95\% CL): in the range 12\%-16\% for $10^{-2}-0.2~\mathrm{M}_\odot$
and 20\% at $0.5~\mathrm{M}_\odot$
for the All sample and down to 10\% and 
below 20\% in the range $10^{-3}-0.2~\mathrm{M}_\odot$
and 26\% at $0.5~\mathrm{M}_\odot$
for the Bright sample.

\subsection{The nature of the observed events} \label{sec:nat}

\emph{What is the nature of the observed lensing systems?}, or,
to rephrase it, \emph{is there any evidence for
a signal from non self-lensing population, namely, from MACHOs?}.
We now attempt to address this issue starting from 
the results presented in the previous section,
and in particular moving beyond the simple
statistics based on the event number
presented in the last section (Fig.~\ref{fig:nexp}).

\subsubsection{The number of the events and their spatial distribution}

OGLE-II reported one candidate event (All sample),
for which we evaluate 0.44 expected self-lensing events. 
OGLE-III reported three (two) candidate events from the All (Bright) sample,
with our evaluation of an expected self-lensing signal of  1.62 (0.93) events,
respectively. Finally, EROS-2 reported one event out
of a selected Bright subsample of sources, 
for which we evaluate an expected self-lensing signal of 0.46 events.
Based on the underlying Poisson nature of the statistics
of the detected events, the observed signal, according to the number of events,
can be therefore fully explained by the expected self-lensing signal,
according to our model most of it coming from faint  SMC stars.
As remarked, assuming an SMC model
in agreement with that in \cite{eros98},
for the same overall mass of the SMC luminous
population, the number of expected self-lensing
events is about twice as large than what we obtain
with our fiducial model. This model, leaving aside
the discussed issue of the spatial distribution
of the events, would then lead to an even
stronger confidence on the reliability of this  outcome.
 
Although the statistics of events is not large,
we may try to move beyond this considerations by
exploiting the additional characteristics of the observed signal.
We have already discussed the spatial distribution
within the framework of the optical depth analysis.
In fact we expect the increase of the SMC self-lensing optical depth
moving towards the SMC centre to be reflected
in a corresponding increase of the expected signal
in terms of the number of events. If we bin the observed field of view
as in Section~\ref{sec:tau}, we indeed find such an increase.
For the MW lens populations, on the other hand,
the expected distribution in terms of number of events
is found to be roughly flat. These results
are not surprising as the bins
are chosen to contain an equal number of sources
and therefore the expected signal follow
the underlying optical depth profiles.
To gain some further insight, we may evaluate
the fraction of expected SMC self-lensing events,
for each experiment, lying outside the contour of equal
expected number of sources fixed
by the position of the reported events.
It results: 34\%, 28\% and 38\% 
for EROS-2, OGLE-II and OGLE-III,
respectively. For OGLE-III the reported value is derived
for the outermost event, and
in this case we may also evaluate the fraction of expected events
lying within the contour of the inner reported event, which turns
out to be of 47\% (the corresponding
fractions of source stars in these four cases
are, respectively, 44\%, 33\%, 58\% and 29\%). 
Assuming the \cite{eros98} model,
we find again that the signatures of a much
stronger gradient moving towards
the SMC centre, namely the fraction,
are significantly smaller (and larger
for the last considered case). About 15\%
of the events, for EROS-2, OGLE-II and OGLE-III, 
are expected out of the contour of equal number
of sources fixed by the position  of the outermost reported event 
(for a fraction of source stars equal
to 38\%, 36\% and 50\%, respectively)
and, for OGLE-III, 72\% of the events
are expected (with 35\% of the source stars)
within the contour of the innermost reported event.

\subsubsection{The duration distribution}

Besides their position, the events are characterized
by the duration. This is a useful statistics
to our purposes as the duration distribution is
independent from the expected event number.
As a test case against the distribution
of the observed durations we consider
the expected distribution for self-lensing events,
SMC self lensing and MW disc lenses, both
stars and brown dwarf (Fig.~\ref{fig:dgte}).
As a result, we find that the duration of two out of five
events falls outside the 16\%-84\% range of probability
for self-lensing lenses.
More specifically, there is only about 5\% probability 
to get a self-lensing event duration
shorter (longer) than that of OGLE-SMC-04 (OGLE-SMC-02).
As apparent also from Fig.~\ref{fig:dgte},
short events are more likely for brown dwarf
lenses, which represent, however,
only about 10\% of the overall expected signal (Table~\ref{tab:nevt}).
On the other hand, very long duration events
look difficult to be explained
(the case of OGLE-SMC-02,
for which additional information is available
to characterize the event, is further discussed below). 
To further quantify these statements, we may attempt to compare
statistically the observed and the expected distributions.
To this purpose, we consider the smaller but homogeneous
set of the three All sample OGLE-III microlensing candidates,
which span, incidentally, the full range
of observed durations, $t_\mathrm{E}=(18.6,\,45.5,\,195.6)~\mathrm{d}$.
To compare the observed and expected self-lensing duration
distribution first we make use of the Kolmogorov-Smirnov test
which allows one to evaluate the probability
of accepting the null hypothesis that the two distributions
are indeed equal. This is known, however, to be specifically sensitive
to compare the \emph{median} values of the distributions,
and in fact we find a rather large probability, $62\%$.
A similar statistics, built to be more sensitive
to the outliers, is that of Anderson-Darling \citep{numrec92}
for which indeed the probability, which
we evaluate through a simulation, drops to $32\%$.

A final remark, quite apparent at glance
from Fig.~\ref{fig:dgte}, is that, if not 
completely by self-lensing events,
the very large spread of the observed durations distribution
makes unlikely the possibility
of explaining all the events by a single mass
MACHO population (\emph{if} any MACHOs).

\subsubsection{The likelihood analysis} \label{sec:like}
\begin{figure}
\includegraphics[width=84mm]{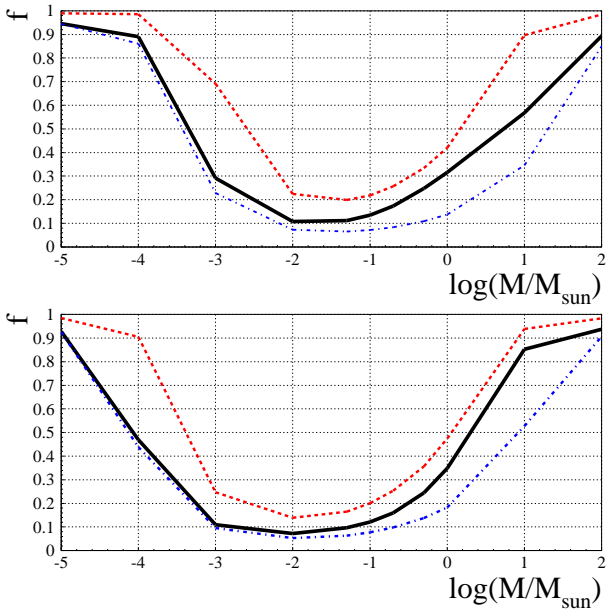}
\caption{Likelihood analysis: 
95\% CL upper limit for the
mass halo fraction in the form 
of MACHO, $f$, as a function
of the MACHO mass (in solar mass units)
for All (OGLE-II and OGLE-III, top panel)
and Bright (OGLE-III and EROS-2) sample of sources, solid lines.
The dashed (dot-dashed) lines indicate the results
we obtain under the hypothesis that the observed
event are due to MACHO lensing (self lensing), respectively.
}
\label{fig:like}
\end{figure}

The likelihood analysis allows us
to further address the issue
of the nature of the events
and in particular to quantify
the limits for the halo mass
fraction in form of compact halo objects, $f$.
To this purpose we proceed as detailed
in Appendix~\ref{app:like}, taking, for reference,
the expression of the likelihood
in terms of the differential rate with respect
to the event duration. This leads 
to include within the analysis both the line-of-sight position 
and the duration of the observed events. Fixing the MACHO mass
as a parameter, given the likelihood,
we may build the probability distribution
for $f$, $P\left(f\right)$, by Bayesian inversion
assuming a constant prior different from zero
in the interval $(0,1)$.

In Fig.~\ref{fig:like}, we show the results
of the likelihood analysis, in particular
we report the 95\% CL upper limit for $f$
as a function of the MACHO mass. The curve shape
reflects in part that of the expected
number of MACHO events reported in Fig.~\ref{fig:nexp},
weighted, however, by the number and the specific characteristics
of the observed events. Here, we consider separately
the two cases of the four reported candidates from the All sample
(OGLE-II and OGLE-III), and the three reported candidates from the Bright sample
(OGLE-III and EROS-2), top and bottom panel in Fig.~\ref{fig:like},
respectively. For the All sample, we find
the lowest constraint for $f$ in the mass
range $10^{-2}-10^{-1}~\mathrm{M}_\odot$,
with $f\le 11-13\%$. The upper limit
then reduces to $30\%$ at $1~\mathrm{M}_\odot$,
at the same level than that at $10^{-3}~\mathrm{M}_\odot$.
Whereas this second result is driven by the small
number of expected MACHO lensing events there,
the increase of the upper limit for $f$ in the
$10^{-1}-1~\mathrm{M}_\odot$ range is rather driven
by the characteristics of the events.
The overall shape behaviour of the $f$ upper limit is similar
moving to the Bright sample. Here, however, thanks to the
enhanced EROS-2 sensitivity to short duration events,
we can put an $f\sim 10\%$ upper limit constraint
over the range $10^{-3}-10^{-1}~\mathrm{M}_\odot$,
with the lowest value at $f\sim 7\%$ for $10^{-2}~\mathrm{M}_\odot$.
The increase in the upper limit above $10^{-1}~\mathrm{M}_\odot$
is somewhat faster in this case, with $f<35\%$ at $1~\mathrm{M}_\odot$.
These behaviours, for the All and Bright sample,
can also be more specifically explained
on the basis of the event characteristics.
In particular, the very short $t_\mathrm{E}=18.60~\mathrm{d}$
OGLE-SMC-04 event, both in the All and in the Bright sample,
somehow drives the results for low mass values,
up to about $10^{-1}~\mathrm{M}_\odot$,
while the two long events, OGLE-SMC-02 and EROS2-SMC-1,
both for the Bright sample, become relevant for
large mass values (this is made apparent in particular
by the, relatively, stronger constraint
on $f$ for $10~\mathrm{M}_\odot$ in the All with
respect to the Bright sample, where in the second case 
two out of three events are very long duration ones).

A better understanding of the likelihood analysis results
comes from the inspection of the dashed and dot-dashed
upper limits in Fig.~\ref{fig:like} where
we report the results of the likelihood
analysis under the assumption
that the events are due to MACHO lensing
(self lensing), respectively
(we remark that these results are based on the 
MACHO lensing signal only, namely the expected self-lensing rate 
does not enter the likelihood function).
Both for the All and the Bright sample
of sources, assuming that the events
are self lensing, the differences
in the upper limit for $f$, comparing
with the solid line where no hypotheses are done
on the lens nature, are small up
to about $10^{-2}~\mathrm{M}_\odot$ and then start increasing
up to a rather large size. This somehow measures the extent
to which, within the likelihood analysis, the events
are weighted as self lensing compared to MACHO lensing.
In particular, this confirms that the MACHO lensing signal
is strongly suppressed especially for low-mass values.
The rather large difference in the two curves
(solid and dot-dashed) for $10~\mathrm{M}_\odot$ 
can also be understood on this basis recalling
the very long durations events present in the All and Bright sample
of sources. The dashed curve built assuming
the events are MACHO, on the other hand, can be looked
at as giving the more conservative upper limit
for $f$, regardless of the characteristics of the events
(just as the dot-dashed discussed above gives
the less conservative one which can be obtained based on the available data).
For the All sample, this is at about 20\% level 
in the range $10^{-2}-10^{-1}~\mathrm{M}_\odot$, reducing
to 40\% for $1~\mathrm{M}_\odot$, whereas for
the Bright sample the limit is up to about, in absolute sense, 6\% smaller
in the lower range and, as before, significantly smaller
at $10^{-3}~\mathrm{M}_\odot$ and, on the other hand,
somewhat larger for $1~\mathrm{M}_\odot$. Overall,
the difference between the two curves (dashed and solid)
is about constant at 10\% (in absolute sense) for the Bright sample
all the way from  MACHO mass above $10^{-3}~\mathrm{M}_\odot$.
For the All sample the difference is also of about 10\%
but only within the range $10^{-2}-1~\mathrm{M}_\odot$.
Below and above these values, at $10^{-3}$ and $10~\mathrm{M}_\odot$,
the shape of the dashed curve then reflects
the drop in the expected number of MACHO lensing events.
The numerical detail of these results,
also distinguishing each experimental set-up,
is reported in Table~\ref{tab:like}.
\begin{table}
\caption{Likelihood analysis: 95\% CL upper limit for $f$, the halo
mass fraction in form of MACHOs  for the All and the Bright sample.
We report the results for the All sample for OGLE-II (1), OGLE-III (2) and OGLE-II plus OGLE-III (3-6);
for the Bright sample for EROS-2 (1), OGLE-III (2) and EROS-2 plus OGLE-III (3-6).
In columns (1-3) the likelihood 
is expressed in terms of the differential
rate with respect to the event duration;
in column (4) the likelihood
is evaluated taking into account
the number of expected events (Appendix ~\ref{app:like});
in column (5), (6) the upper limit on $f$
is evaluated under the hypothesis
that the observed events are (not) MACHOs.
}
\begin{center}
\begin{tabular}[h]{ccccccc}
Mass & (1) & (2) & (3) & (4) & (5) & (6)\\
\hline
All sample &&&&&&\\
$10^{-3}~\mathrm{M}_\odot$ & 0.904 & 0.326 & 0.290 & 0.516 & 0.691 & 0.228\\ 
$10^{-2}~\mathrm{M}_\odot$ & 0.665 & 0.120 & 0.107 & 0.167 & 0.224 & 0.073\\
$0.1~\mathrm{M}_\odot$     & 0.579 & 0.152 & 0.135 & 0.167 & 0.218 & 0.071\\
$0.5~\mathrm{M}_\odot$     & 0.789 & 0.261 & 0.247 & 0.255 & 0.333 & 0.109\\
$1~\mathrm{M}_\odot$       & 0.861 & 0.325 & 0.315 & 0.321 &  0.420 & 0.137\\
$10~\mathrm{M}_\odot$      & 0.928 & 0.615 & 0.567 & 0.755 & 0.896 & 0.345\\
\hline
Bright sample &&&&&&\\
$10^{-3}~\mathrm{M}_\odot$ & 0.133 & 0.437 & 0.109 & 0.186 & 0.247 & 0.095\\
$10^{-2}~\mathrm{M}_\odot$ & 0.105 & 0.160 & 0.072 & 0.111 & 0.139 & 0.053 \\
$0.1~\mathrm{M}_\odot$     & 0.204 & 0.202 & 0.120 & 0.162 & 0.200 & 0.076\\
$0.5~\mathrm{M}_\odot$     & 0.468 & 0.366 & 0.245 & 0.290 & 0.356 & 0.138 \\  
$1~\mathrm{M}_\odot$       & 0.670 & 0.478 & 0.348 & 0.386 & 0.474 & 0.182\\
$10~\mathrm{M}_\odot$      & 0.944 & 0.859 & 0.851 & 0.873 & 0.938 & 0.526 \\
\hline
\end{tabular}
\end{center}
\label{tab:like}
\end{table}

We can compare the upper limits on $f$ obtained within
the likelihood analysis to those derived from
the Poisson upper limits discussed in Section~\ref{sec:res}
and Fig.~\ref{fig:nexp}. Overall, they appear,
quite significantly at least in a relative sense,
\emph{larger}.  The driving motivation
is the characterization as \emph{indistinguishable}
of the ``signal'' with respect to the
underlying ``background'' one assumes
to evaluate the upper limits, for the signal, 
for the Poisson distribution with a background \citep{FC98}.
The degeneracy in the lensing parameter space
justify to some extent this characterization;
however, the likelihood analysis allows one
to take advantage of the specific characteristics
of the observed events. It is also 
interesting to note that, within the scheme
of the \cite{FC98} statistics, assuming
the mean expected background to be 
equal to the observed signal, one would get
to about equal (and for some values of the MACHO mass,
even tighter) constraints for $f$.

The results discussed above on the likelihood
are obtained with our fiducial model.
We may wonder what happens when using the 
\cite{eros98} model for which
we expect about twice as much self-lensing events.
As a result, we find that the upper limit we obtain
in this case are indeed somewhat smaller,
but the difference turns out never to exceed,
in absolute sense, 3\%-4\%,
namely, the two results are about equal.
The underlying reason can be traced back
in the spatial distribution of the observed events
as compared to the expected signal which is,
for the SMC self-lensing component, 
extremely more clustered around the SMC centre
in the \cite{eros98} model. From a methodological
point of view, this outcome clearly highlights
the extent to which it is relevant
to include within the analysis all the information
available to draw  meaningful conclusions on the
MACHO lensing as compared to the self-lensing one,
which is specifically what is made possible
by the likelihood analysis.

To conclude on the likelihood analysis
we may address the question of whether
the result we obtain is biased by our choice
of expressing the likelihood in terms
of the differential rate rather than
considering the number of expected events (Appendix~\ref{app:like}).
In this second case, corresponding
to the line of sight of the observed events,
the MACHO lensing and the self-lensing signals
are compared based on the number of 
expected events only (in particular with no reference,
therefore, to  the observed event duration).
As a result, Table~\ref{tab:like}, the upper limit in this case turns out
to be larger, and this can be understood on the basis
that in most cases the observed durations
are in agreement with the self-lensing expected
ones and, if not, as for the very long duration
events, the expected MACHO lensing signal
is however quite small: overall this attributes
to the self-lensing rate more weight with respect
to the MACHO lensing one than in the case
where only the expected number (along
a specific line of sight) is considered.
The overall change, however,
turns out to be not too large,
in an absolute if not relative sense,
compared to the results
reported in Fig.~\ref{fig:like}.
In particular, the difference is about
5\% and up to 10\% in the mass range
$10^{-2}-1~\mathrm{M}_\odot$ and 
$10^{-3}-10~\mathrm{M}_\odot$
for the All and the Bright sample,
respectively, with the difference
which tends to be larger for small
mass values. The larger difference,
about 20\%, we find for the All sample
at $10^{-3}$ and $10~\mathrm{M}_\odot$
should, as above, be traced back
in the drop of the expected
MACHO lensing signal to a level almost
compatible with the self-lensing one.

\subsubsection{The projected velocity: the case of OGLE-SMC-02}

\begin{figure}
\includegraphics[width=84mm]{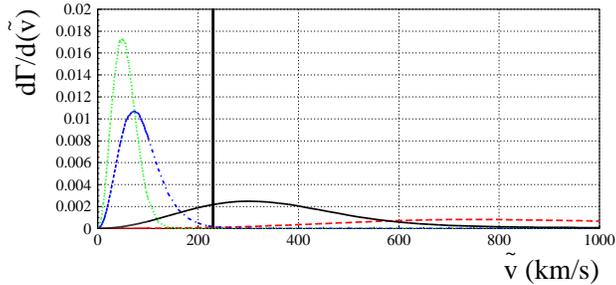}
\caption{Normalized differential rate 
$\mathrm{d}\Gamma/\mathrm{d}\tilde{v}$,
where $\tilde{v}=v/(1-D_l/D_s)$ is the projected velocity,
along the line of sight of OGLE-SMC-02
microlensing candidate event ($t_\mathrm{E}=195.6~\mathrm{d}$).
In particular, we show the result
for different lens populations:
MW thin disc, MW thick disc, SMC self lensing and
MW MACHO lensing (dotted, dash-dotted, dashed, solid lines,
respectively). The vertical solid line
represents the estimated observed value.
}
\label{fig:vtilde}
\end{figure}

Although the likelihood analysis is driven
by the characteristics of the observed events,
it remains a statistical approach on a full
set of events. Further insight into the nature
of the lenses can be gained for those
events for which additional information
is available. Within the present set of events,
this is specifically the case for the long duration
($t_\mathrm{E}=195.6~\mathrm{d}$) OGLE-SMC-02 candidate
event. As discussed in Section~\ref{sec:obs},
\cite{dong07} did conclude strongly in favour
of the MW MACHO nature of this event.
In general, the analysis of the MACHO hypothesis suffers from
the degeneracy within the lensing parameter
space of the unknown lens mass. As apparent also from inspection
of the differential rate distribution, Table~\ref{tab:te} and Fig.~\ref{fig:dgte},
a long-duration event as OGLE-SMC-02
might indeed be explained by a heavy ($\sim (1-10)~\mathrm{M}_\odot$) MACHO,
possibly a black hole. 
This conclusion is however dependent on an hypothesis on the lens mass.
Within their analysis, \cite{dong07} 
could get rid of this limitation.
In particular, they were able
to provide an estimate of the projected velocity \citep{gould94}
$\tilde{v}=v/(1-D_l/D_s)$, a quantity which
is only weakly dependent on the lens mass (and altogether
independent if assuming a delta mass function),
with $\tilde{v}_\mathrm{obs}\sim 230~\mathrm{km~s}^{-1}$.
We have evaluated the expected differential distribution for
$\mathrm{d}\Gamma/\mathrm{d}\tilde{v}$
given our fiducial model for the SMC self-lensing,
MW disc and MW MACHO populations
(from Eq.~\ref{eq:rate}, we evaluate
$\mathrm{d}\Gamma/\mathrm{d}\tilde{v} = 
\mathrm{d}\Gamma/\mathrm{d}v \times (1-D_l/D_s)$).
The result of this analysis is reported in Fig.~\ref{fig:vtilde}.
At glance, the estimated observed value, $\tilde{v}_\mathrm{obs}$, is 
in good agreement with the MW MACHO lensing population projected velocity distribution
and at odds with that of the self-lensing populations.
In particular, the probability of getting
a smaller (larger) value than  $\tilde{v}_\mathrm{obs}$
is below 1\% for SMC star (MW disc) lenses.

Looking back to the initial issue, the nature
of the observed events, we have therefore
shown that, although the bare number 
of observed versus expected events may suggest
that all of them may be attributed
to self-lensing populations, a more thorough analysis,
considering the full set of available
event characteristics, leads us  at very least
to soften this conclusion. On the other hand,
if any, the extent of the MACHO population compatible
with the available data would clearly remain very small.

\section{Comparison to previous analyses} \label{sec:compare}

We can compare our results with those
presented by the OGLE collaboration, 
in particular for the OGLE-III SMC All sample data set \citep{ogle11b}.
Overall, the upper limits for $f$ we obtain
are tighter (in particular we estimate an increase,
with respect to their results,
of the expected number of MACHO events
driving the upper limit statistics
of about 60\%); however, we get to only
partly understand the underlying
reason of this disagreement.
OGLE base their statistical analysis on the 
following approximated estimate
of the expected number of (Galactic) MACHO lensing events 
\citep{eros07} (we recall in particular
the underlying assumption $<{\cal E}>={\cal E}(<t_\mathrm{E}>)$
which becomes less and less accurate moving to small values
of the MACHO mass for which the resulting differential rate,
$\mathrm{d}\Gamma/\mathrm{d}t_\mathrm{E}$
is significantly different from zero just where
the efficiency is negligible)
\begin{equation} \label{eq:nexptau}
N_\mathrm{exp} = \frac{2}{\pi}\,N_\mathrm{obs} T_\mathrm{obs}\,\tau\,
\frac{{\cal E}(<t_\mathrm{E}>)}{<t_\mathrm{E}>}\,.
\end{equation}
OGLE then derives the upper limit on $f$
following the recipe of \cite{FC98}
for a Poisson statistics with a background
as that we have carried out in Section~\ref{sec:res}
with the assumption that the mean
value of the expected background signal
(self lensing) is equal to the
observed number of events. 
As discussed in Section~\ref{sec:like},
comparing to the likelihood, 
this analysis tends to give
more conservative upper limits whenever
the two analyses are carried out
with coherent values of the expected 
self-lensing signal. The hypothesis
of OGLE, which in this case
overestimates the expected background, 
drives however the limits much closer to the ones we obtain
with the likelihood analysis.
Here comes the second  caveat,
regarding Eq.~\ref{eq:nexptau},
as OGLE uses for $\tau$ and 
$<t_\mathrm{E}>$ values from previous
analyses carried out towards the LMC,
specifically $\tau=4.7\times 10^{-7}$,
for a full MACHO halo, with the Einstein time scaling with
the lens mass as $<t_\mathrm{E}>=70\,\sqrt{m}$ \citep{ogle10,ogle11b}.
According to our analysis the corresponding, average,
values should read instead $\tau=6.3\times 10^{-7}$,
for a full MACHO halo towards the SMC,
with $<t_\mathrm{E}>=66\,\sqrt{m}$,
which overall makes a relative increase,
for the expected number of events, of about 40\%. 
For reference we also note that, when considering 
the line of sight towards
the LMC, coherently with these values towards the SMC,
one should rather use  $\tau=4.5\times 10^{-7}$
and $<t_\mathrm{E}>=62\,\sqrt{m}$ \citep{novati09b,novati11}.
As for the LMC, the difference from previous values
follows from the assumed distance to the Galactic
Centre, $8~\mathrm{kpc}$ rather than $8.5~\mathrm{kpc}$,
and from the inclusion, within the likelihood analysis,
of the components of the bulk motion of the relative velocity. 

It is also interesting to compare
the line of sight towards the SMC to that towards the LMC.
Here again we take advantage of the OGLE-III analysis \citep{ogle11},
which we also have discussed in \cite{novati11},
to carry out an homogeneous comparison.
As for the MACHO lensing signal
it is useful again to start  from
Eq.~\ref{eq:nexptau}.
The LMC counts  almost four times more numerous 
source stars, which more than compensates the decrease
in the ratio $\tau/<t_\mathrm{E}>$ discussed above.
Based on these terms, fixed $T_\mathrm{obs}$,
one should expect almost three times more MACHO
lensing events towards the LMC than towards the SMC.
The efficiency however, at least following the
OGLE-III analysis, tends to reduce this difference,
especially for the Bright sample
(indeed, the efficiency towards the LMC
greatly varies from field to field
according to the relative crowding,
whereas towards the, less crowded SMC fields, it results 
roughly constant across the monitored fields of view
and, especially for the Bright sample, relatively larger).
As for self lensing, lensing by SMC stars
is strongly enhanced by the SMC morphology
compared to LMC self lensing;
on the other hand the overall MW disc 
lensing signal is relatively much
more important for the LMC because of the much
larger extension of the monitored field of view.
Overall, the expected self lensing signal towards the LMC
turns out to be about twice as large as that towards the SMC.
Face to these changes in the expected signal,
the observed rate, for the OGLE-III analysis,
turns out, with the caveat of the small statistics,
to be fully compatible towards the two lines of sight:
for the All sample, two (three) candidate events are reported 
towards the LMC (SMC). These effects therefore combine so that the constraints
on $f$ from the LMC turns out to be tighter.

\section{Conclusions} \label{sec:end}

In this paper, we have discussed
the results of the microlensing campaigns carried out
towards the SMC by the EROS \citep{eros07} and the OGLE
collaborations \citep{ogle10,ogle11b}. 
In particular, we have addressed the issue of the nature
of the lens of the observed events, 
either to be attributed to ``self lensing'',
where the lens belong to some luminous component
(either of the SMC or of the MW disc)
as opposed to MACHO lensing from 
the putative population of 
dark matter compact halo objects of the MW.
To this purpose, we have carried
out analyses of the microlensing optical depth and
of the expected signal
based on the evaluation of the microlensing rate.

Overall, five microlensing candidates have been
reported (one each by EROS-2 and OGLE-II and three by OGLE-III).
Whereas in terms of number of events, this may be fully explained
by the expected self-lensing signal
(out of which about 90\% is expected from SMC self lensing),
the analyses  based on the event characteristics,
line-of-sight position and duration,
and for one event on the evaluation of the projected velocity,
rather suggest that not all the events may be attributed
to this lens population. In particular,
2.1 (1.4) self lensing events are expected,
to be compared to 4 (3) observed events,
depending on the sample of sources considered.
Two events (both reported by OGLE-III) have durations
lying outside the 95\% limits of the expected self-lensing signal
(one shorter, one longer). The long-duration event is the same
for which the projected velocity analysis,
which strongly favour a non-self lensing nature
of the lens, has been carried out \citep{dong07}. 
Additionally, we have discussed the spatial distribution  
of the observed events as compared to the
profile of the SMC self lensing optical depth.
Finally, both the event line-of-sight position and duration 
enter the likelihood analysis.
Taking into account the expected signal
of the self-lensing and MACHO lensing populations,
this allows us to quantify the resulting upper limit on the halo mass fraction
in form of MACHOs, $f$. In particular, it results that the upper
limit at 95\% CL is lowest, about 10\%, at $10^{-2}~\mathrm{M}_\odot$,
and then reduces to above 20\% for $0.5~\mathrm{M}_\odot$ MACHOs.
Overall, these limits are somewhat less tight than
those obtained by analogous analyses
carried out towards the LMC (\citealt{eros07,ogle11}) where, also compared
to a somewhat larger expected signal in terms
of MACHO lensing events, the number 
of observed events is not correspondingly larger.
Larger set of events, hopefully
available in the next future thanks to the ongoing OGLE-IV
and MOA-II campaigns should
provide further insight in this problem. 

The expected SMC self-lensing signal is driven
by the underlying model of the SMC luminous components
for which, in this work, we have taken advantage of several
recent analyses (\cite{bekki09b,subsub12,grebel12} and references therein), which however
still do not provide a full coherent picture
of its formation history, dynamic and morphology. Among the more relevant quantities
for microlensing purposes, the value for the 
line-of-sight depth seem quite well established.
There remain however still open questions
as the overall luminous SMC mass and the exact balance
between the old and the young star populations.
A correct model for the SMC luminous 
remains a key issue for the understanding 
of the microlensing signal.
Indeed, a larger set of events would make even more important
a detailed knowledge of the SMC morphology,
providing a further relevant tool of analysis
to address the issue of the lens nature.

\section*{Acknowledgements}
We thank the referee for valuable comments and remarks.
We thank {\L}ukas~Wyrzykowski  for several
fruitful discussions on the OGLE analysis.
We thank Kenji~Bekki for discussions on the SMC
model. SCN and GS acknowledges support by
the Swiss National Science Foundation during part of this work.
PJ and SM acknowledge support by the Swiss
National Science Foundation.

\bibliographystyle{mn2e}

\bibliography{biblio}

\appendix
\section{The likelihood analysis} \label{app:like}

The observation of microlensing events
follows a Poisson distribution with the expected
number determined according to the given model.
Suppose we have $N_\mathrm{obs}$ observed events 
for an expected signal of $N_\mathrm{exp}$  events.
Introducing a binning of the parameter space
which specifies the model, we can write down
the joint probability distribution for
obtaining $N_\mathrm{obs}$ events, 
namely the likelihood, as
the product over  the $N_\mathrm{bin}$ bins to have 
$n_i$ observed events for an expected signal $x_i$,
with $x_i$ being the parameter for the Poisson
distribution in each separate bin . For a suitable
choice of the binning, we can then make $n_i$ equal 
either to 0 or to 1, namely we can get to infinitesimal
bins so to have either none or one event per bin \citep{gould03},
which is the second step in Eq.~\ref{eq:like},
whereas in the last step one makes use of the fact
that the extent of the bin where no events
are observed is indeed overall infinitesimal
\begin{eqnarray} \label{eq:like}
L &= &\prod_{i=1}^{N_\mathrm{bin}} \frac{\exp(-x_i) x_i^{n_i}}{n_i!} =
\prod_{i=1}^{N_\mathrm{obs}} \exp(-x_i)\,x_i \!\!\!\!\!\!
\prod_{i\notin (1,N_\mathrm{obs})}\!\!\!\!\!\!\exp(-x_i)\nonumber\\
& = &\exp(-N_\mathrm{exp}) \prod_{i=1}^{N_\mathrm{obs}} x_i\,.
\end{eqnarray}
In the last term, the product runs over the bins containing only one observed event.
Out of the likelihood, given the prior distribution and by Bayesian inversion, we 
can build the probability distribution for the parameters
of interest. In the following,
we consider the likelihood as a function of $f$,
the halo mass fraction in form of MACHOs, keeping
the MACHO mass fixed as a parameter.

The terms $x_i$, being related to the expected number of events
per bin, are proportional to the microlensing rate. As
a possible approach, one can introduce a binning
in the duration, $\Delta t_\mathrm{E}$, and then
reduce to Eq.~\ref{eq:like} by the limit $\Delta t_\mathrm{E}\to 0$.
In this case $x = \mathrm{d}\Gamma/\mathrm{d}t_\mathrm{E}$,
evaluated at the value of the observed durations,
$t_\mathrm{E,obs}$. This is the likelihood expression used, for
instance, in the analyses of the MACHO group (\cite{macho00}
and references therein). Alternatively,
one can directly consider $x$ as the number of expected events per bin evaluated
according to Eq.~\ref{eq:nexp}. 
This gives the likelihood analysis used,
for instance, within the analysis of M31 pixel lensing results of
the POINT-AGAPE collaboration \citep{novati05}. Whatever the choice,
the underlying structure of Eq.~\ref{eq:like} drives
the resulting limit on $f$.

The rate, and therefore the expected number of events,
can be looked at as the sum of two terms: the self-lensing
contribution plus the MACHO lensing contribution modulated the
multiplicative factor $f$. As a first remark, we note that
in the exponential term, $\exp(-N_\mathrm{exp})$, the number
of expected self-lensing events drops out as a constant.
In particular, this implies that for no observed events,
either assuming that the observed events are due
to self lensing the resulting limits on $f$ are independent from 
the expected self-lensing signal and are driven
by the expected number of MACHO lensing events only.
In the more general case of $N_\mathrm{obs}>0$, the exponential decrease 
of $f$ is modulated by $N_\mathrm{obs}$ factors of the kind
$a+f b$, where $a$ and $b$ are constants with respect to $f$
and linked to the expected self-lensing and MACHO lensing signal,
respectively. To the purpose of the evaluation of the probability
distribution for $f$, $P(f)$, only the ratio $b/a$ matters so that
whatever factor coming in front of both
of them drops out in the normalization of $P(f)$.
In particular, for the choice mentioned above,
$x = \mathrm{d}\Gamma/\mathrm{d}t_\mathrm{E}$,
when calculating the differential rate at the
observed duration value, the efficiency term 
${\cal{E}}(t_\mathrm{E})$ does drop out
(being usually given as a unique function for all the lens populations considered). 
On the other hand, when considering for $x$ the number of expected events,
the efficiency ${\cal{E}}(t_\mathrm{E})$ enters
in an essential way, whereas the constants that drop out are the number of sources
and the overall time span of the experiment,
so that in particular one can consider, for instance,
as the infinitesimal bin choice, the lines of sight corresponding
to each observed event. 
It is also important to keep track that in these two cases
one is in fact weighting the event characteristics
in a different way. In the first case, both the event line-of-sight position
and the duration enter the likelihood (with the relevant caveat
that the duration is not modulated by the detection efficiency). 
In the second, the results are driven specifically
by the expected number of events within the chosen bins,
namely the line of sight position. The outcome
is therefore expected to be more similar
to the analysis carried out based
on the number of events according to the
Poisson distribution. This is not surprisingly
as the underlying statistics is the same,
with the important caveat, however,
that within this likelihood-based analysis
also the observed event spatial distribution
is included within the analysis. 

In the present analysis, we consider the joint results
from more than one experiment. In this case,
the probabilities, and therefore the different likelihood terms, multiply.
Each experiment is characterized by his own number
of expected events and this fixes,
through the exponential term in the likelihood,
the relative weight of each of them.
On the other hand, all the reported events,
appearing in the product, enter the likelihood
on the same footing.

\label{lastpage}

\end{document}